\documentclass[
floatfix, noshowpacs, preprintnumbers, twocolumn, amsmath, amssymb, aps, prl, superscriptaddress]{revtex4-1}

\usepackage{times}
\usepackage{bm}
\usepackage{amsmath}
\usepackage{amssymb}
\usepackage{graphicx}
\usepackage{xcolor}
\usepackage{bbold}
\usepackage{url}
\usepackage{graphicx}
\usepackage[caption=false]{subfig}
\usepackage{amsthm}
\usepackage{soul}
\usepackage{framed}
\usepackage[utf8]{inputenc}
\usepackage{bbold}
\usepackage{wrapfig}
\usepackage{multirow}
\usepackage{booktabs}

\newcounter{suppEq}
\begin{document}

\title{Optimal photonic indistinguishability tests in multimode networks}

\author{Niko Viggianiello}
\affiliation{Dipartimento di Fisica, Sapienza Universit\`{a} di Roma, Piazzale Aldo Moro 5, I-00185 Roma, Italy}

\author{Fulvio Flamini}
\affiliation{Dipartimento di Fisica, Sapienza Universit\`{a} di Roma, Piazzale Aldo Moro 5, I-00185 Roma, Italy}

\author{Marco Bentivegna}
\affiliation{Dipartimento di Fisica, Sapienza Universit\`{a} di Roma, Piazzale Aldo Moro 5, I-00185 Roma, Italy}

\author{Nicol\'o Spagnolo}
\affiliation{Dipartimento di Fisica, Sapienza Universit\`{a} di Roma, Piazzale Aldo Moro 5, I-00185 Roma, Italy}

\author{Andrea Crespi}
\affiliation{Istituto di Fotonica e Nanotecnologie, Consiglio Nazionale delle Ricerche (IFN-CNR), 
	Piazza Leonardo da Vinci, 32, I-20133 Milano, Italy}
\affiliation{Dipartimento di Fisica, Politecnico di Milano, Piazza Leonardo da Vinci, 32, I-20133 Milano, Italy}

\author{Daniel J. Brod}
\affiliation{Perimeter Institute for Theoretical Physics, 31 Caroline Street North, Waterloo, ON N2L 2Y5, Canada}
\affiliation{Instituto de F\'isica, Universidade Federal Fluminense, Av. Gal. Milton Tavares de Souza s/n, Niter\'oi, RJ, 24210-340, Brazil}
\author{Ernesto F. Galv\~{a}o}
\affiliation{Instituto de F\'isica, Universidade Federal Fluminense, Av. Gal. Milton Tavares de Souza s/n, Niter\'oi, RJ, 24210-340, Brazil}

\author{Roberto Osellame}
\affiliation{Istituto di Fotonica e Nanotecnologie, Consiglio Nazionale delle Ricerche (IFN-CNR), 
	Piazza Leonardo da Vinci, 32, I-20133 Milano, Italy}
\affiliation{Dipartimento di Fisica, Politecnico di Milano, Piazza Leonardo da Vinci, 32, I-20133 Milano, Italy}

\author{Fabio Sciarrino}
\email{fabio.sciarrino@uniroma1.it}
\affiliation{Dipartimento di Fisica, Sapienza Universit\`{a} di Roma, Piazzale Aldo Moro 5, I-00185 Roma, Italy}

\begin{abstract}
\textbf{Particle indistinguishability is at the heart of quantum statistics that regulates fundamental phenomena such as the electronic band structure of solids, Bose-Einstein condensation and superconductivity. Moreover, it is necessary in practical applications such as linear optical quantum computation and simulation, in particular for Boson Sampling devices. It is thus crucial to develop tools to certify genuine multiphoton interference between multiple sources. Here we show that so-called Sylvester interferometers are near-optimal for the task of discriminating the behaviors of distinguishable and indistinguishable photons. We report the first implementations of integrated Sylvester interferometers with 4 and 8 modes with an efficient, scalable and reliable 3D-architecture. We perform two-photon interference experiments capable of identifying indistinguishable photon behaviour with a Bayesian approach using very small data sets. Furthermore, we employ experimentally this new device for the assessment of scattershot Boson Sampling. These results open the way to the application of Sylvester interferometers for the optimal assessment of multiphoton interference experiments.}
\end{abstract}


\maketitle

\section{Introduction}

\begin{figure}[htbp]
	\centering
	\includegraphics[width=\linewidth]{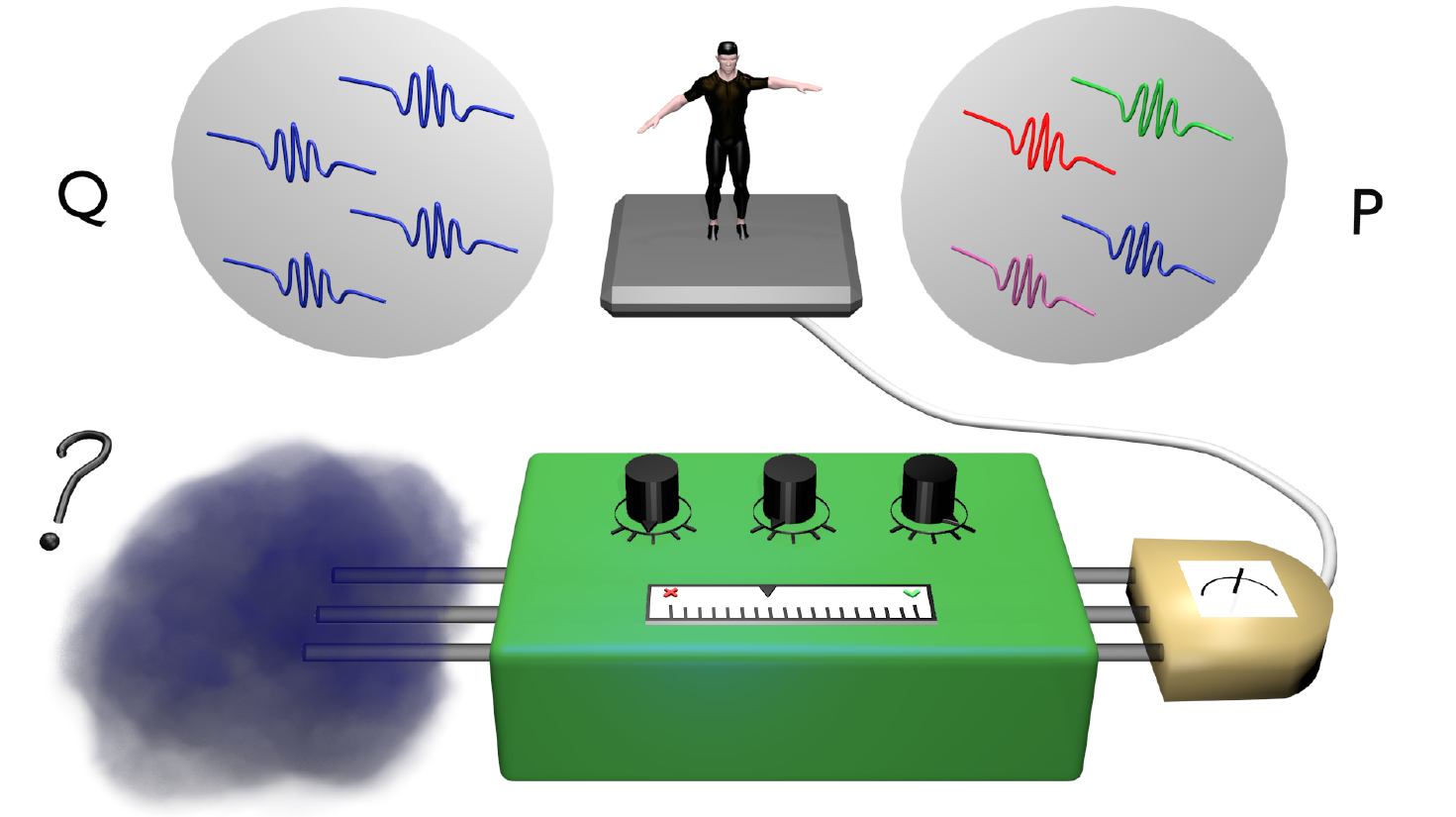}
	\caption{Searching for the optimal indistinguishability test. An unknown multiphoton state is injected into a multimode interferometer, designed to discriminate between the two hypotheses: either the photons are indistinguishable ($Q$) or  distinguishable ($P$). The internal parameters can then be tuned until the most efficient test is found, for which the distributions corresponding to the two hypotheses $P$ and $Q$ are as different as possible, as measured by the total variation distance of the output distributions.} 
	\label{Concept}
\end{figure}

The observation that fundamental particles may be intrinsically indistinguishable is a counter-intuitive feature of quantum mechanics. As a consequence, their collective behaviour is governed by quantum statistics, and results in important phenomena such as Bose-Einstein condensation and the electronic band structure of solids. Moreover, many-body bosonic interference is also at the very heart of quantum photonic computation \cite{KLM,nielsen_chuang,Aaronson10}. 
In this context, Boson Sampling devices harness multiphoton interference effects to provide evidence of a superior quantum computational power with current state-of-the-art technology \cite{Aaronson10, Broome13, Spring13, Tillmann12, Crespi12,Spagnolo13,Carolan14, Spagnolo14,Carolan15,Bentivegna15,Loredo16,Wang17,He_Pan17}.
The task of certifying genuine many-boson interference is thus expected to find numerous applications in photonic quantum information, for validating the functioning of Boson Sampling experiments 
\cite{Spagnolo13, Aaronson14, Bentivegna14, Carolan14, Spagnolo14, Tichy14, Carolan15,Crespi16,Liu16,Walschaers16,Bentivegna16,Shch16,Wang16}
and, more generally, as a diagnostic tool for quantum optical devices \cite{Aolita15,Spring16}. 
A well-known approach to experimentally test the degree of indistinguishability between photons involves the use of the Hong-Ou-Mandel effect 
\cite{Hong87}, 
for which various generalizations have been proposed to account for multiple photon sources 
\cite{Oua88, Walborn03, Sagioro04,Spagnolo13trit,Till15,Mens16,Agne16}. 
A natural problem that arises is then to determine which interferometers allow for an optimal assessment of the indistinguishability of multiphoton states (see Fig. \ref{Concept}).

In this research article we introduce a new approach for the task of discriminating distinguishable and indistinguishable photon behaviour in multimode photonic networks, and perform an experimental demonstration with two-photon inputs in laser-written integrated devices with an optimal interferometer design. We consider a set-up in which probabilistic single-photon sources are coupled to different sets of input ports of an interferometer, as required for instance in scattershot Boson Sampling experiments 
\cite{Lund14, Bentivegna15,Latmiral16}. 
To assess the optimality of the interferometer we maximize the total variation distance (TVD) between the output probability distributions corresponding to indistinguishable and distinguishable photons, showing that in certain scenarios the optimal solution is provided by so-called Sylvester interferometers. Finally, we perform a scattershot Boson Sampling experiment in the 4-mode Sylvester device, and verify the capability of the developed approach to validate the collected data with a small number of events.

\section{Sylvester matrices for optimal indistinguishability tests}

A linear interferometer is a device that implements a linear map between input and output creation operators, namely: $a^{\dagger}_{in_k} \to \sum_j U_{j,k}\, a^{\dagger}_{out_j}$. A $m$-mode interferometer is then completely specified by its corresponding $m \times m$ unitary matrix $U$, which can be decomposed into two-mode beam splitters and single-mode phase shifters \cite{Reck94}. The evolution of a $n$-photon state is ruled by the unitary transformation and by the degree of indistinguishability of the input particles. Let $P=\{p_i\}$ ($Q=\{q_i\}$) denote the output probability distribution of an interference experiment when the input photons, all injected in different input modes, are perfectly distinguishable (indistinguishable) and the occupation number is detected at each output port. 
The total variation distance (TVD) is then defined as
\begin{equation}
T(P,Q) = \frac{1}{2} \sum_i |p_i - q_i| .
\label{TVDformula}
\end{equation}
The TVD corresponds to the highest possible difference in probabilities that $P$ and $Q$ can assign to the same event. It is also closely related to hypothesis testing, since $1 - T(P,Q)$ is a lower bound on the sum of probabilities of false positive and false negative results \cite{lehmann_romano}. More practically, a larger value of $T(P,Q)$ means that a smaller set of experimental data is sufficient to discriminate between the hypotheses that one is sampling from $P$ or $Q$ up to a given level of confidence.

\begin{figure*}[htbp]
	\centering
	\includegraphics[width=\linewidth]{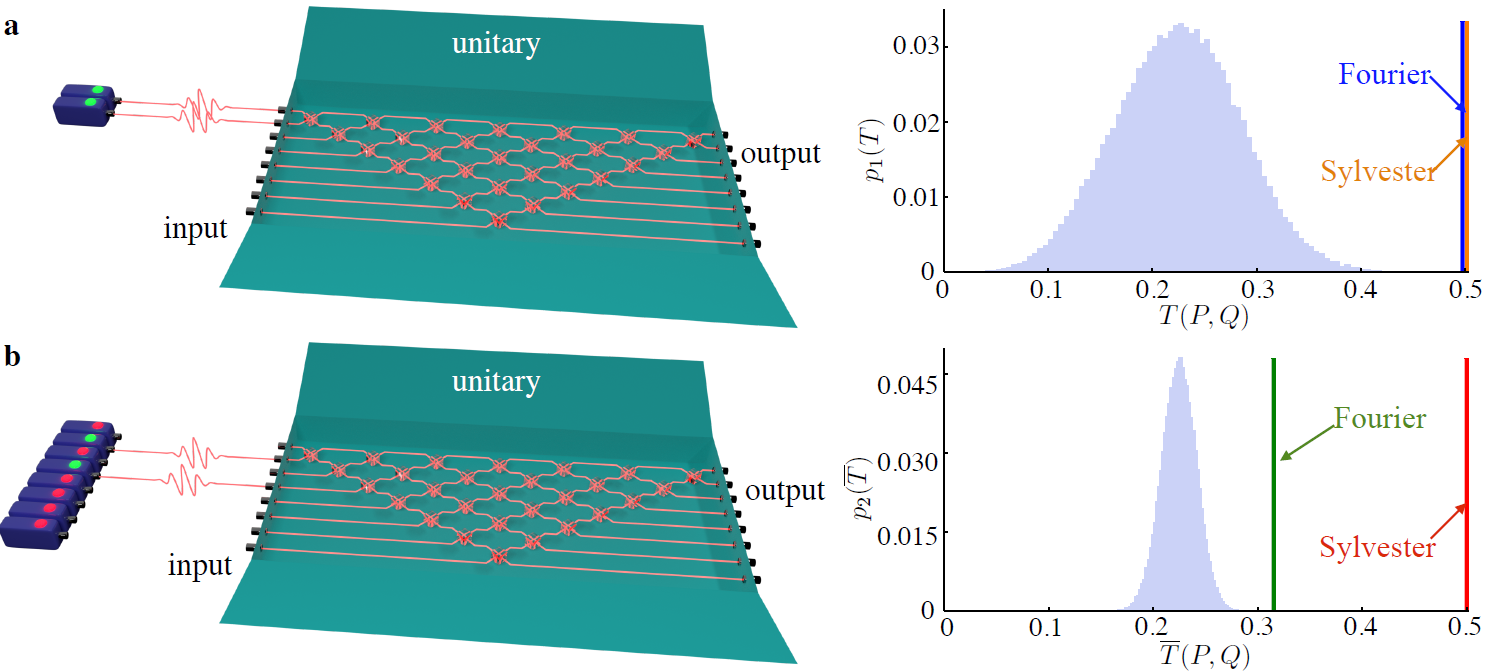}
	\caption{ Optimal tests for fixed-input and scattershot configurations. {\bf a}, Left figure: fixed input configuration in the $n=2$, $m=8$ case, where two fixed input modes are injected with single photons. Right figure: estimate of the distribution $p_{1}(T)$ of the total variation distance $T(P,Q)$ for uniformly random unitaries in this scenario, obtained numerically by sampling over $10^5$ unitaries. For specific choices of input modes, both Fourier (blue vertical bar) and Sylvester (orange vertical bar) interferometers are optimal.
		{\bf b}, Left figure: scattershot configuration in the $n=2$, $m=8$ case, where different sets of two input modes are picked at random. Right figure: estimate of the distribution $p_{2}(\overline{T})$ of the average total variation distance $\overline{T}(P,Q)$ for uniformly random unitaries in this scenario, obtained numerically by sampling over $10^5$ unitaries. In the scattershot case the Sylvester transformation (red vertical bar) is optimal, while the Fourier is not.}
	\label{Concept2}
\end{figure*}

We consider two scenarios depending on how the input photons enter the interferometer: (i) the input ports are fixed (see Fig. \ref{Concept2}a) and (ii) the input combination is randomly chosen for each event (see Fig. \ref{Concept2}b). This second situation is typical of more sophisticated interferometric experiments with multiple probabilistic sources, such as scattershot Boson Sampling experiments \cite{Lund14, Bentivegna15}. To search for the optimal platform, a natural symmetry we can impose on the interferometer is that the output probability for single photons should be uniform for all input ports. Mathematically, this corresponds to a unitary matrix $U$ with complex entries of modulus $1/d$, known as a Hadamard matrix. A specific subcase is provided by Sylvester transformations \cite{Crespi15}, described by $m=2^p$-dimensional unitary matrices defined by the recursive rule:
\begin{equation}
S(2^p)=\left(
\begin{array}{cc}
S(2^{p-1}) & S(2^{p-1}) \\
S(2^{p-1}) & -S(2^{p-1})
\end{array}
\right)
\end{equation}
with $S(2^0)=S(1)=(1)$ and $p$ any positive integer. Another notable example of Hadamard matrix is the Fourier one:

\begin{equation}
(U_{m}^{\operatorname{F}})_{l,q}=\frac{1}{\sqrt{m}}\, e^{i\, \frac{2 \pi l q}{m}}.
\label{fourier:formula}
\end{equation}

\noindent which will be another subject of our analysis. For two-photon experiments, we can apply \eqref{TVDformula} to check that the Sylvester transformation results, in the case of interferometers with $m=4$ modes, in a $T_{\mathrm{S}}=0.5$ independently of which pair of input modes is used, and thus on average (over different pairs of inputs) $\overline{T}_{\mathrm{S}}=0.5$. As we show later, the Sylvester interferometer is optimal for all possible choices of two-photon input configurations. The Fourier interferometer results instead in a lower average $\overline{T}_{\mathrm{F}}=0.333$, reaching the highest value $T_{\mathrm{F}}=0.5$ only for cyclic inputs, i.e. $n$-photon Fock states in interferometers with $n^p$ modes, where the occupied modes $j_{r}^s$ are given by $j_{r}^s=s+(r-1) \, n^{p-1}$, with  $r=1, \ldots, n$ and $s=1, \ldots, \,n^{p-1}$.  For two-photon experiments in interferometers with $m=8$ modes, among all known Hadamard matrices the Sylvester transformation is again the one with the largest average TVD $\overline{T}_{\mathrm{S}}=0.5$ ($T_{\mathrm{S}}=0.5$, and thus optimal, for all pairs of inputs). The Fourier interferometer reaches ${T}_{\mathrm{F}}=0.5$ for cyclic inputs, while on average we obtain only $\overline{T}_{\mathrm{F}}=0.3153$ due to inputs with lower TVD. Even if we consider general unitary matrices, for two-photon experiments in $m=4$ and $m=8$ modes a random sampling of $10^{5}$ matrices of each size fails to find matrices outperforming the Sylvester matrices, which provides evidence for the optimality of Sylvester interferometers. In the $n=2$ $m=4$ case, the optimality can be proved also via a numerical maximization of the TVD over the parameters of general 4-mode interferometers.

Interestingly, we found that for scenarios with different number of photons $n$ and modes $m$,  the optimal interferometers may not even be Hadamard. As an example, for $n=3$, $m=4$, the best design we found corresponds to the unitary matrix
\begin{equation}
U_4= \frac{1}{2} \left(
\begin{array}{cccc}
1 & 1 & \sqrt{2} & 0 \\
1 & -1 & 0 & \sqrt{2} \\
1 & 1 & - \sqrt{2} & 0 \\
1 & -1 & 0 & -\sqrt{2} \\
\end{array}
\right),
\end{equation}
with average (over all collision-free inputs) $\overline{T}_{U_4}=0.53125$, beating for example the Fourier and Sylvester interferometers, which result in $\overline{T}=0.3125$. In Supplement 1 we list the best interferometers we found numerically for a few different scenarios.

\begin{figure*}[htbp]
	\centering
	\includegraphics[width=\linewidth]{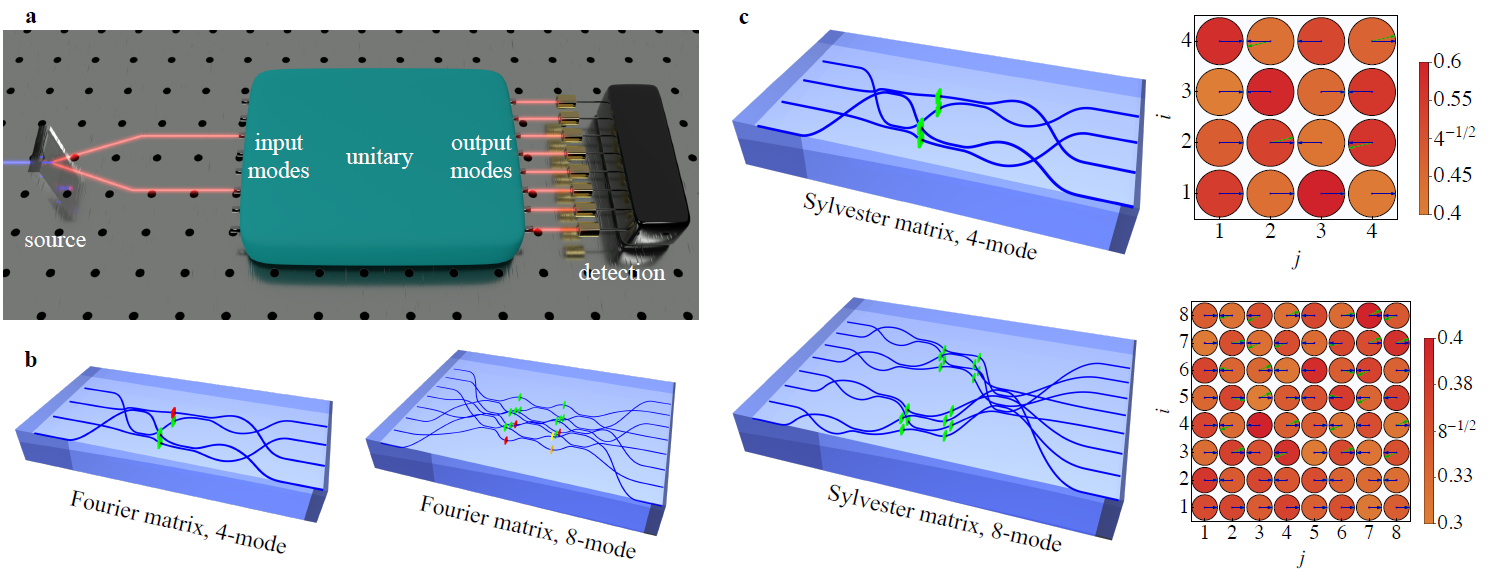}
	\caption{\textbf{a,} Scheme of the experimental approach for the optimal test. A photon source based on parametric down-conversion produces pairs of photons that can be injected in any possible input combination of a unitary transformation, with photodetection at the output. \textbf{b,} Internal structure for the 4- and 8-mode laser-written Fourier interferometers. \textbf{c,} Internal structures for the 4-mode and 8-mode implemented Sylvester interferometers, and corresponding representation of the tomographically reconstructed matrices: disk color indicates the modulus of each matrix element, with an arrow representing its phase (green: reconstructed unitary, blue: theoretical matrix). In the circuital structures of panels (b) and (c), colored boxes correspond to phase shifters. Green boxes: $\phi=0$. Red boxes: $\phi=\pi/2$. Orange boxes: $\phi=3 \pi/4$. Yellow boxes: $\phi=\pi/4$.}
	\label{setup}
\end{figure*}

Here we report the experimental demonstration of optimal distinguishability tests, which we carried out using two custom-designed integrated photonic circuits fabricated on a glass substrate via the femtosecond laser writing technology \cite{Osellame03,DellaValle09,Marshall09}. A possible approach to the implementation of Fourier and Sylvester matrices on photonic platforms could exploit the decompositions of Refs. \cite{Reck94,Clements16}, which allow to implement arbitrary unitary operations through an in-plane cascade of beam-splitters and phase shifters but are sensitive against losses inside the device. A more effective approach for the implementation of this class of interferometers is found by adopting an efficient, scalable and reliable three-dimensional interferometer design enabled by the interaction between non-first neighbor modes arranged in a suitable lattice \cite{Barak07} (Fig. \ref{setup}). This architecture presents several advantages with respect to the conventional decomposition in beam splitters and phase shifters \cite{Reck94,Clements16}. Indeed, this approach is efficient since it allows to significantly reduce the number of optical elements from $O(m^{2})$ to $O(m\, \log \, m)$. Furthermore, as shown in Supplement 1, it is scalable to a larger number of modes. Finally, it is also reliable since the depth of the circuit is small $[O(\log m)]$ and the layout is symmetric with respect to the input modes, thus being intrinsically more robust against internal losses than the decomposition of Ref. \cite{Reck94}. These features permit to achieve higher fidelities in the implemented devices.

To probe the device for the reconstruction of its internal operation, we injected the interferometers with one- and two-photon states produced via a spontaneous parametric downconversion process \cite{Burnham70}. The tomographic reconstruction of the processes exploits a priori knowledge of the internal structure, minimizing a suitable $\chi^2$ function with respect to the unknown internal phases and transmissivities \cite{Crespi16}. In Supplement 1 we provide more details about the design and the tomographic reconstruction. All tests have been carried out also on Fourier interferometers of equivalent dimensions ($m=4,8$), to provide a comparison between the two performances and highlight the advantages of the new approach. For more details on the implementation of the Fourier transforms refer to Ref. \cite{Crespi16}.

\section{Measurement of the TVD}

The TVD between distributions obtained from distinguishable and indistinguishable photons has been estimated experimentally for two-photon experiments in 4-mode and 8-mode Sylvester and Fourier interferometers. The two condictions are reached by properly introducing a relative temporal difference between the two photons using delay lines (see Supplement 1). As previously discussed, a distinctive feature of Sylvester interferometers is that the TVD has the same value for two-photon experiments using any pair of input ports. This value is never outperformed by Fourier interferometers, being matched only for cyclic inputs.

\begin{figure*}[htbp!]
	\centering
	\includegraphics[width=\linewidth]{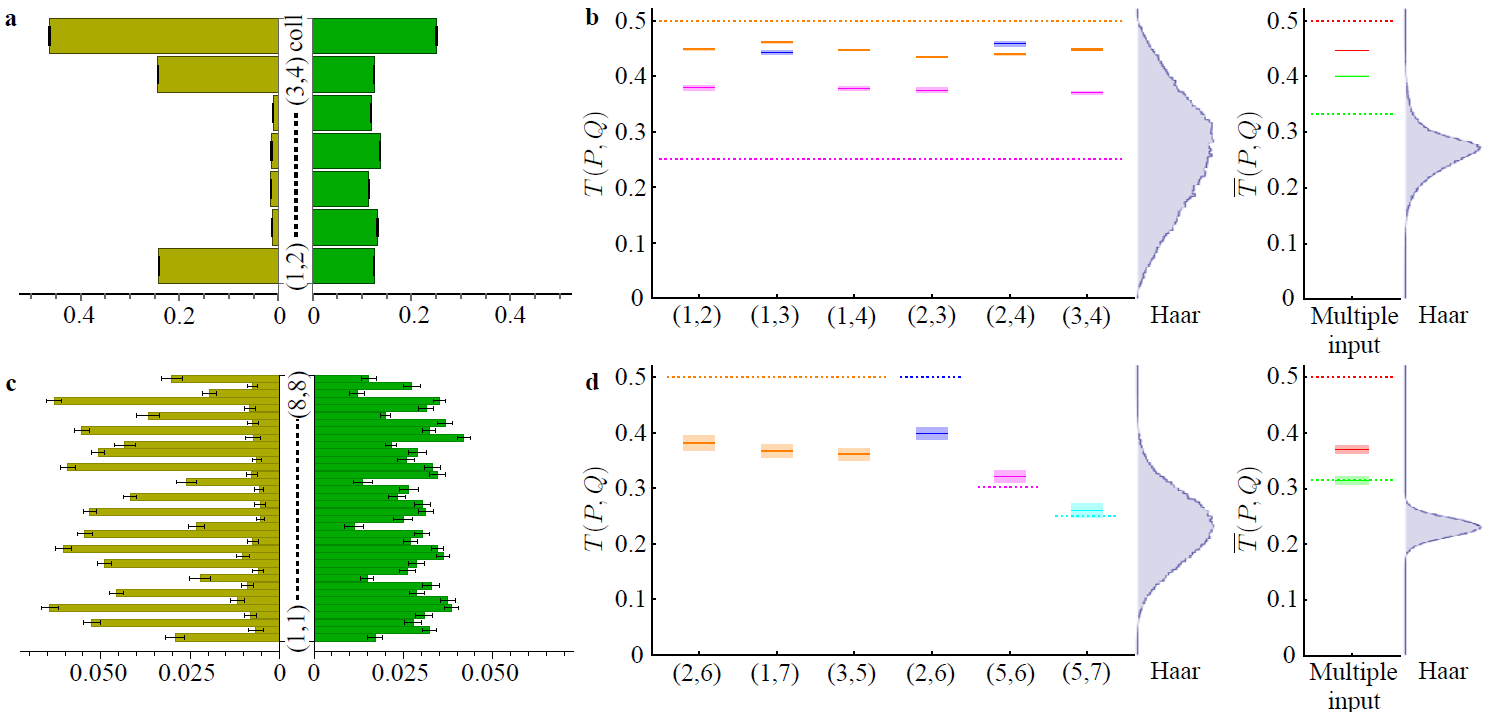}
	\caption{ Experimental data from Sylvester interferometers and comparison with Fourier ones. Two-photon experimental data from 4-mode and 8-mode Sylvester interferometer. {\bf a,c}, Output probability distributions for ({\bf a}) input (1,2) of the 4-mode Sylvester interferometer and ({\bf c}) input (1,5) of the 8-mode interferometer. Each plot compares the distributions obtained from indistinguishable (dark yellow) and distinguishable (green) input photons. The additional bar \textit{coll.} in ({\bf a}) takes into account all collision events. {\bf b,d}, Total variation distances between the measured output distributions for indistinguishable and distinguishable photons in the 4-mode ({\bf b}) and 8-mode ({\bf d}) interferometers, plotted for each fixed input (left) and for the multiple input configuration (right). On each plot, the distributions of $T(P,Q)$ and $\overline{T}(P,Q)$ for Haar-random transformations (obtained by sampling $10^5$ unitaries) are shown. Legend - Rectangles: 1$\sigma$ regions for the experimental data. Dashed lines: theoretical predictions. For $m=4$, the expected values are obtained by using the extra collision bin. Red: Sylvester matrix, multiple input. Green: Fourier matrix, multiple input. Orange: Sylvester, fixed input. Blue: Fourier, fixed cyclic inputs. Purple, Cyan: Fourier, fixed non-cyclic inputs.}	
	\label{test}
\end{figure*}

In the case of the 4-mode Sylvester and Fourier interferometers, all $\binom{4}{2}=6$ possible pairs of different input ports have been injected, measuring for each pair all the collision-free output distributions (that is, events with at most one photon per output port). Additionally, we included an extra bin in the distribution accounting for the probability of collision events (see Fig. \ref{test}a). For distinguishable photons, this collision probability can be calculated using the measured probabilities for single-photon experiments. For indistinguishable photons, the additional information provided by the measured two-photon Hong-Ou-Mandel visibilities, together with the single-photon data, allows for an estimate of the collision probability. The adoption of this extra bin prevents the appearance of pathological unitaries in our numerical search, with high values of TVD for the collision-free subspace but unacceptably small probability of collision-free events, which strongly reduce the amount of detected signal (see Supplement 1 for more details). Adjusting the distinguishability of the input photons we could then estimate the TVD according to \eqref{TVDformula}. The experimental results are reported in Fig. \ref{test}b. In the fixed input case, the Sylvester interferometer (for all inputs) and the Fourier one (only for cyclic inputs) reaches similar values within the range $0.435 \leq T^{(4)} \leq 0.461$. When averaging over all inputs, the 4-mode Sylvester interferometer results in $\overline{T}^{(4)}_{\mathrm{S}}=0.4465 \pm 0.0006$, thus clearly outperforming the Fourier interferometer that yields $\overline{T}^{(4)}_{\mathrm{F}}=0.4003 \pm 0.0016$. 
In the case of the 8-mode interferometers, all two-photon output events were measured, including those with collisions (see Fig. \ref{test}c). We probed the interferometers with three different pairs of input ports, corresponding to the three classes of input pairs that, for the Fourier interferometer, result in different TVDs. Again, the TVD for the Fourier interferometer reaches the same value of the Sylvester only for cyclic input pairs. The measured average values were $\overline{T}^{(8)}_{\mathrm{F}}=0.314 \pm 0.007$ and $\overline{T}^{(8)}_{\mathrm{S}}=0.370 \pm 0.007$, where the average $\overline{T}^{(8)}_F$ was estimated by weighting the three representative input states by the multiplicity of the corresponding classes with different TVDs (Fig. \ref{test}d). Curiously, the measured $\overline{T}^{(4)}_{\mathrm{F}}$ and $\overline{T}^{(8)}_F$ are higher than what was expected theoretically. Indeed, as we show in Supplement 1, a particular feature of the Fourier interferometer design is that manufacturing errors result in interferometers which are closer to the Sylvester one, thus increasing the TVD. We performed the same analysis also with $10^5$ random unitaries using the Haar-uniform measure. In Fig. 4b-d (right) it is possible to notice how the mean values of TVDs are lower than both Fourier and Hadamard unitaries. Our results are compatible with the theoretical prediction that Sylvester interferometers provide optimal discrimination between distinguishable and indistinguishable photons in all scenarios (fixed and multiple input), outperforming, in particular, Fourier interferometers in the multiple input configuration.

\section{Bayesian hypothesis test}

\begin{figure*}[htbp!]
	\centering
	\includegraphics[width=0.9\linewidth]{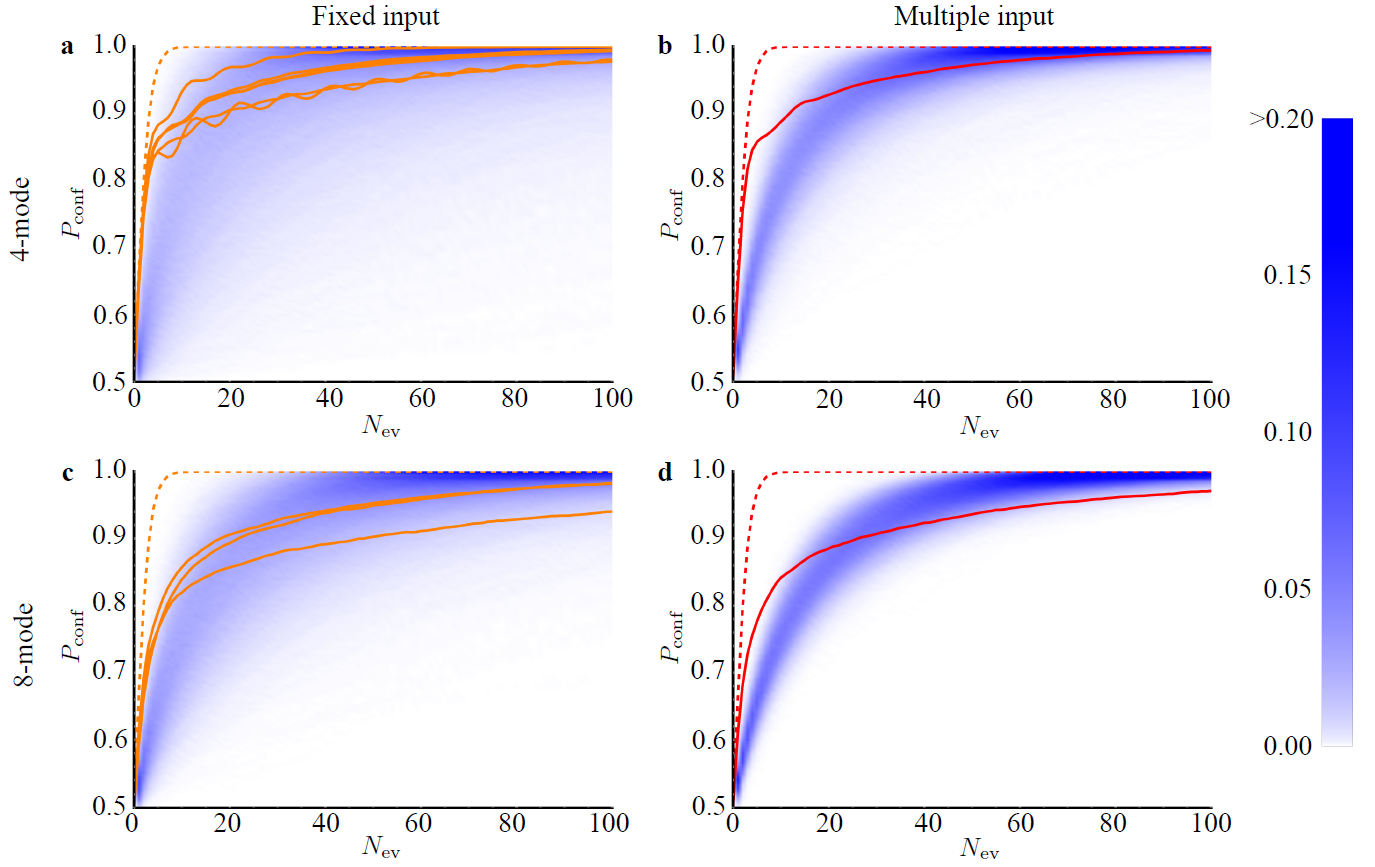}
	\caption{ Bayesian simulation of hypothesis testing. Probability $P_{\mathrm{conf}}$ of correctly identifying the type of data sample (distinguishable or indistinguishable photons) in a Bayesian test, as a function of sample size (number of events), for the Sylvester interferometer.
		{\bf a}, 4-mode interferometer, fixed input states. {\bf b}, 4-mode interferometer, multiple input configuration. {\bf c}, 8-mode interferometer, fixed input states. {\bf d}, 8-mode interferometer, multiple input configuration. For the 8-mode devices, inputs used were (2,6), (1,7), (3,5). Dashed curves: theoretical predictions. Solid curves: average of $10^4$ distinct samples generated numerically from the experimentally measured distributions. Red curves: Sylvester matrix, multiple input configuration. Orange curves: Sylvester matrix, different input states. Blue regions: contour plots obtained from a numerical simulation over $10^4$ Haar-random matrices. Simulated data for the Haar-random case include partial photon indistinguishability $x=0.95$ equal to the experimental data, characterized by performing Hong-Ou-Mandel interference in a 50/50 beam-splitter.}	
	\label{bayesianTest}
\end{figure*}

A larger TVD should allow us to discriminate more readily the two hypotheses (distinguishable or indistinguishable photons). To illustrate this feature, in Fig. \ref{bayesianTest} we show the results of a Bayesian analysis with data samples numerically generated from the measured distributions, compared with those expected from ideal interferometers (see Ref. \cite{Bentivegna14} and Supplement 1 for more details on the Bayesian validation test). For a given sample size, we use a likelihood ratio test \cite{Bentivegna14} to update a prior which initially assigns equal probabilities to the two hypotheses $P$ and $Q$, thus assuming no a priori knowledge. To verify the optimality of Sylvester interferometers, the results obtained are compared with a sample of $10^4$ Haar-random unitaries. The figure of merit is the confidence probability $P_{\mathrm{conf}}$ that a data sample is assigned to the corresponding correct hypothesis.

\begin{figure*}[htbp!]
	\centering
	\includegraphics[width=1\linewidth]{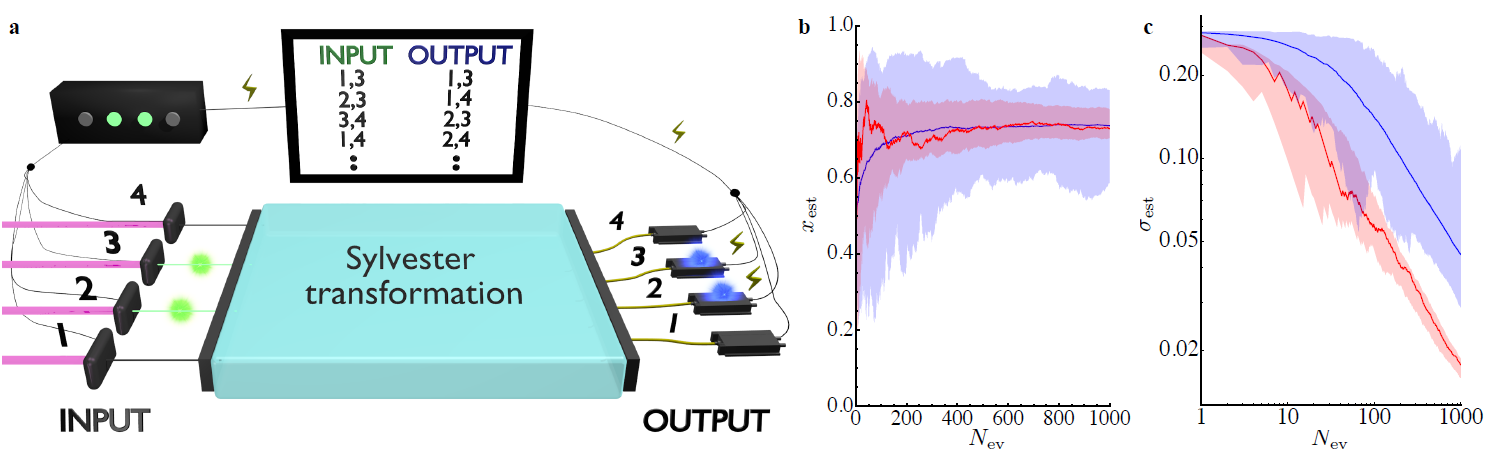}
	\caption{ Bayesian inference on experimental scattershot boson sampling. Inference of the photon indistinguishability $x$ as a function of sample size (number of events) from two-photon scattershot Boson Sampling experimental data in a 4-mode Sylvester chip. {\bf a}, Experimental scheme of the scattershot approach. We use 4-PDC sources to probabilistically inject a different input state, which is retrieved from heralded 4-fold coincidences (two trigger and two output detectors). {\bf b}, Evolution of the estimated value $x_{\mathrm{est}}$ as a function of the number of events and {\bf c}, error $\sigma_{\mathrm{est}}$ inferred on the estimated value $x_{\mathrm{est}}$ from the Bayesian posterior distribution with the 4-mode interferometer (experimental data) as a function of the number of events $N_{\mathrm{ev}}$, compared with Haar-random matrices (numerical simulations). Red solid line: single data sequence for the Sylvester interferometer. Red shaded region: interval included within $M=100$ different data sequences generated by random permutations of the experimentally measured data. Blue solid line: average value obtained by numerically sampled $M=100$ Haar-random unitary. Blue shaded region: interval obtained by numerically sampling $M=100$ Haar-random matrices. The value of $x$ adopted for the Haar-random numerical simulations is equal to the final value $\tilde{x}_{\mathrm{est}}$ estimated from the full set of experimental data.} 
	\label{bayesian_scatter}
\end{figure*}

Note that the dispersion of $P_{\mathrm{conf}}$ for Haar matrices in the fixed input case is larger than in the multiple input configuration. Numerical simulations performed on the Bayesian test show that in presence of partial photon indistinguishability, quantified by a parameter $x \in [0;1]$, the photons to pass the test if $x > 0.788$ for the 4-mode interferometer and $x > 0.685$ for the 8-mode case (see Supplement 1).

\section{Experimental assessment of scattershot Boson Sampling}

Scattershot Boson Sampling has been shown \cite{Bentivegna15} to enable scalable Boson Sampling experiments using probabilistic single-photon sources, at the cost of using a larger number of single-photon sources at the input. To show our approach works well for this multiple-input configuration, we have performed a scattershot Boson Sampling experiment by connecting each input mode of our 4-mode Sylvester interferometer to an independent heralded parametric down-conversion source (see Fig. \ref{bayesian_scatter}a). This single setup simultaneously samples all 6 different two-photon input states. We have collected data both for the indistinguishable and distinguishable photon cases, obtained by appropriately adjusting the relative time delay between the input paths. We have then performed a step forward with respect to the binary Bayesian test applied previously (that is, evaluating which is the most likely hypothesis between $Q$ and $P$). More specifically, we consider a different scenario where the aim is to assess the output data by estimating the value of the mutual indistinguishability of the PDC sources in the scattershot experiment. The set of hypotheses to be tested is now provided by a convex combination of indistinguishable $Q$ (distinguishable $P$) photons $H(x) = x Q + (1-x) P$. As previosly introduced, $x \in [0;1]$ is a continuous real parameter quantifying the degree of indistinguishability between the photons generated by different sources. We have then applied Bayesian inference \cite{Pezz2006,Svore2014} to the measured experimental data to estimate the value of $x$, starting from a uniform prior $\mathcal{P}(x)$ and subsequently updating the distribution according to the Bayes rule after the observation of a given data sample (see Supplement 1 for more details on this approach).

Figs. \ref{bayesian_scatter}b-c show the results of the Bayesian inference performed by using the measured $\sim 17000$ experimental scattershot data samples (in this case, no extra bin for collision events has been included). The final estimated value $\tilde{x}_{\mathrm{est}}$ of the indistinguishability parameter $x$, obtained by using the complete set of experimental data, is $\tilde{x}_{\mathrm{est}} = 0.738 \pm 0.004$. Such value is compatible with the one obtained from a characterization with Hong-Ou-Mandel interference in a 50/50 beam-splitter $x_{\mathrm{HOM}} = 0.79 \pm 0.06$, and with the estimated interval obtained with an alternative method based on binary likelihood ratio tests $x_{\mathrm{LR}} \in [0.734;0.742]$ (see Supplement 1).  

The performance obtained with the experimental data on Sylvester interferometer are then compared with the one achievable with Haar random transformations, showing the optimality of the Sylvester matrix. Indeed, the value of the estimation error $\sigma_{\mathrm{est}}$ is lower with respect to the Haar ensemble (for the same number of measured events $N_{\mathrm{ev}}$). This confirms that Sylvester interferometers represent a promising platform for the assessment of scattershot Boson Sampling experiments. 

\section{Discussion}

In summary, in this Research Article we propose a method to search for optimal interferometer designs to ascertain photon indistinguishability in multiparticle interference experiments, showing that in certain scenarios Sylvester transformations offer an optimal solution for all input configurations. We then experimentally implement these interferometers for the first time by exploiting a novel 3D-architecture with the femtosecond laser-writing technology. We verify the superior performance of Sylvester interferometers with a Bayesian hypothesis test which correctly identifies whether photons are distinguishable or not using very small data sets. Furthermore, we perform a complete scattershot Boson Sampling experiment with the implemented Sylvester device, showing the capability of identifying the collected data in this more complex multiple input configuration. Several perspectives can be envisaged starting from these results. 

On the experimental side, we show that the optimality of Sylvester interferometers suggests an immediate application in the assessment of scattershot Boson Sampling, that enables scalable experiments even with probabilistic single-photon sources. Furthermore, this class of transformations can represent a promising platform for near-future investigations on deterministic single photon sources \cite{Ding16,Pascale16,Wang16(2),White16}, in order to fully characterize multiphoton interference \cite{Loredo16,Wang17,He_Pan17,Mens16}, and as a diagnostic tool in reconfigurable photonic devices \cite{Carolan15,Harris17}.

On the theoretical side, the identification of optimal interferometer designs for different interferometer sizes and number of photons is a non trivial and interesting open problem. Different scenarios occur in the fixed input and in the scattershot cases. Based on numerical simulations we can extrapolate that for a single input state, the best unitaries can be obtained by embedding a $n \times n$ unitary matrix in a larger $m \times m$ one, acting as the identity on the unoccupied input modes. Conversely, such operational approach does not hold in the multiple input case, where finding the best unitary requires a global optimization. Hence, while Sylvester matrices are optimal for certain sizes, further investigations are necessary to understand the general structures underlying the optimal unitaries for all $(n,m)$.

\section*{Fundings}
ERC-Starting Grant 3D-QUEST (3D-Quantum Integrated Optical Simulation; grant agreement no. 307783): \href{http://www.3dquest.eu}{3D-Quest}; H2020-FETPROACT-2014 Grant QUCHIP (Quantum Simulation on a Photonic Chip; grant agreement no. 641039): \href{http://www.quchip.eu}{QUCHIP}; Brazilian National Institute for Science and Technology of Quantum Information (INCT-IQ/CNPq). This research was supported in part by Perimeter Institute for Theoretical Physics.


\end{document}


\title{Optimal photonic indistinguishability tests in multimode networks: supplementary material}

\author{Niko Viggianiello}
\affiliation{Dipartimento di Fisica, Sapienza Universit\`{a} di Roma, Piazzale Aldo Moro 5, I-00185 Roma, Italy}

\author{Fulvio Flamini}
\affiliation{Dipartimento di Fisica, Sapienza Universit\`{a} di Roma, Piazzale Aldo Moro 5, I-00185 Roma, Italy}

\author{Marco Bentivegna}
\affiliation{Dipartimento di Fisica, Sapienza Universit\`{a} di Roma, Piazzale Aldo Moro 5, I-00185 Roma, Italy}

\author{Nicol\'o Spagnolo}
\affiliation{Dipartimento di Fisica, Sapienza Universit\`{a} di Roma, Piazzale Aldo Moro 5, I-00185 Roma, Italy}

\author{Andrea Crespi}
\affiliation{Istituto di Fotonica e Nanotecnologie, Consiglio Nazionale delle Ricerche (IFN-CNR), 
	Piazza Leonardo da Vinci, 32, I-20133 Milano, Italy}
\affiliation{Dipartimento di Fisica, Politecnico di Milano, Piazza Leonardo da Vinci, 32, I-20133 Milano, Italy}

\author{Daniel J. Brod}
\affiliation{Perimeter Institute for Theoretical Physics, 31 Caroline Street North, Waterloo, ON N2L 2Y5, Canada}
\affiliation{Instituto de F\'isica, Universidade Federal Fluminense, Av. Gal. Milton Tavares de Souza s/n, Niter\'oi, RJ, 24210-340, Brazil}
\author{Ernesto F. Galv\~{a}o}
\affiliation{Instituto de F\'isica, Universidade Federal Fluminense, Av. Gal. Milton Tavares de Souza s/n, Niter\'oi, RJ, 24210-340, Brazil}

\author{Roberto Osellame}
\affiliation{Istituto di Fotonica e Nanotecnologie, Consiglio Nazionale delle Ricerche (IFN-CNR), 
	Piazza Leonardo da Vinci, 32, I-20133 Milano, Italy}
\affiliation{Dipartimento di Fisica, Politecnico di Milano, Piazza Leonardo da Vinci, 32, I-20133 Milano, Italy}

\author{Fabio Sciarrino}
\email{fabio.sciarrino@uniroma1.it}
\affiliation{Dipartimento di Fisica, Sapienza Universit\`{a} di Roma, Piazzale Aldo Moro 5, I-00185 Roma, Italy}

\maketitle

\section{Optimal distinguishability for different $n$ and $m$}
\label{supp_sec1}

In this section we report results on the optimal interferometer designs we found for the task of testing the hypotheses of distinguishable vs.\ indistinguishable photons, as we increase the number of photons $n$ and modes $m$. As in the main text, the figure of merit is the total variation distance between the two distributions. In Table \ref{tab:tvd}, we report the TVD for combinations of two possible choices. The first is whether we only consider a single input, in which case we choose that input which displays the greater TVD (in order to eliminate ambiguities related to relabeling of the modes), or the average TVD over all possible inputs, which might be more useful for a Scattershot implementation, where one wants to test several different probabilistic sources at once. The second choice is whether we allow collision outcomes (i.e.\ where multiple photons exit in the same mode) or not. In the latter case, one might wish to compute the TVD between the normalized distributions restricted to coincidence outcomes, but that poses a problem: in some cases (e.g.\ in the Hong-Ou-Mandel effect) there are {\bf no} collision-free outcomes, and this means that the renormalized distributions, and thus the TVD, are not well-defined. This has the undesired side-effect of creating pathologies in the numerical search. In order to deal with this, for the collision-free situation we actually bin all collision events as a single ``no-detection'' event in the probability distribution before computing the TVD. 

For each combination of these choices and of values of $n$ and $m$, we also sampled 10000 Haar-random matrices (except $\{m,n\}$ equal to $\{3,8\}$ and $\{2,16\}$, where we sampled 2500, and $\{4,8\}$, where we sampled 100 matrices, due to computational constraints) in order to search for non-Hadamard interferometers that performed better than Sylvester or Fourier matrices. Often the best Haar-random matrix suggested an analytical optimal which we obtained by inspection. These are represented by $U_i$ in the table and discussed below.

{\renewcommand{\arraystretch}{1.2}
	\begin{suppTab}
		\begin{center}
			\resizebox{\columnwidth}{!}{
			\begin{tabular}{  c  c  c  c  c  c  c  }
				\hline
				\; $m$ \; & \; $n$ \;  & $U$ & \; Col, Max \; & \; No-col, Max \; & \; Col, Avg \; & \; No-col, Avg \; \\
				\hline
				\multirow{1}{*}{2} & \multirow{1}{*}{2} & $F_2$ & 0.5 & 0.5 & 0.5 & 0.5 \\ \hline
				\multirow{3}{*}{3} & \multirow{3}{*}{2} & $F_3$ & 0.3333 & 0.3333 & 0.3333 & 0.3333 \\ 
				& & $U_1$ & {\bf 0.5} & {\bf 0.5} & 0.3333 & 0.3333 \\ 
				& & $U_2$ & 0.3951 & 0.3951 & {\bf 0.3951} & {\bf 0.3951} \\ \hline
				\multirow{2}{*}{3} & \multirow{2}{*}{3} & $F_3$ & {\bf 0.6667} & 0.1111 & {\bf 0.6667} & 0.1111 \\ 
				& & $U_3$ & 0.5 & {\bf 0.5} & 0.5 & {\bf 0.5} \\ \hline
				\multirow{10}{*}{4}& \multirow{2}{*}{2} & $S_4$ & {\bf 0.5} & {\bf 0.5} & {\bf 0.5} & {\bf 0.5} \\
				& & $F_4$ & {\bf 0.5} & {\bf 0.5} & 0.3333 & 0.3333 \\ \cline{2-7}
				& \multirow{5}{*}{3} & $S_4$ & 0.3125 & 0.125 &  0.3125 & 0.125 \\ 
				&  & $F_4$ & 0.3125 & 0.125 &  0.3125 & 0.125 \\
				&  & $U_4$ & 0.5625 & {\bf 0.5} &  {\bf 0.5313} & 0.375 \\ 
				&  & $U_5$ & {\bf 0.6667} & 0.3333 &  0.4167 & 0.2778 \\ 
				&  & $U_6$ & 0.5 & {\bf 0.5} &  0.5 & {\bf 0.5} \\ \cline{2-7}
				& \multirow{3}{*}{4} & $S_4$ & {\bf 0.75} & 0.1563 &  {\bf 0.75} & 0.1563 \\
				&  & $F_4$ & {\bf 0.75} & 0.0938 &  {\bf 0.75} & 0.0938 \\
				&  & $U_7$ & 0.5 & {\bf 0.5} &  0.5 & {\bf 0.5} \\ \hline
				\multirow{12}{*}{8}& \multirow{2}{*}{2} & $S_8$ & {\bf 0.5} & {\bf 0.5} & {\bf 0.5} & {\bf 0.5} \\
				& & $F_8$ & {\bf 0.5} & {\bf 0.5} & 0.3153 & 0.3153 \\ \cline{2-7}
				& \multirow{5}{*}{3} & $S_8$ & 0.3125 & 0.2188 &  0.3125 & 0.2188 \\ 
				&  & $F_8$ & 0.4112 & 0.2813 &  0.3407 & 0.2545 \\ 
				&  & $U_8$ & {\bf 0.6667} & 0.3333 &0.1012 & 0.0912 \\
				&  & $U_9$ & 0.5 & {\bf 0.5} &  0.0536 & 0.0536 \\ 
				&  & $U$\textsuperscript{a} & 0.4893 & 0.4311 &  {\bf 0.3658} & {\bf 0.3087} \\ \cline{2-7}
				& \multirow{4}{*}{4} & $S_8$ & {\bf 0.75} & 0.2813 &  {\bf 0.5375} & 0.275 \\ 
				&  & $F_8$ & {\bf 0.75} & 0.3047 & 0.4337 & 0.2501 \\ 
				&  & $U_{10}$ & {\bf 0.75} & {\bf 0.5} & 0.2536 & 0.2013 \\
				&  & $U$\textsuperscript{a} & 0.5359 & 0.3388 & 0.4293 & {\bf 0.2776} \\ \hline
				\multirow{2}{*}{16}& \multirow{2}{*}{2} & $S_{16}$ & {\bf 0.5} & {\bf 0.5} & {\bf 0.5} & {\bf 0.5} \\
				&  & $F_{16}$ & {\bf 0.5} & {\bf 0.5} & 0.3147 & 0.3147  \\ \hline
			\end{tabular}
		}
			\\[10pt]
		\end{center} 
		\caption{TVD between distinguishable and indistinguishable photon distributions. Columns indicates the setting: Max (highest TVD over inputs) or Avg (TVD averaged over inputs), and Col (collision outcomes are considered) and No-col (all collision outcomes are gathered into a single no-click event). $S_n$ and $F_n$ are the Sylvester and Fourier matrices of size $n$, resp., and $U_i$ are other noteworthy matrices as described in the text. Bold numbers correspond to the highest values in that setting and $\{n,m\}$ pair.}
		\begin{flushleft}
			\textsuperscript{a} For these sizes, it was not always possible to identify the best matrices by inspection, so each column might represent a different Haar-random matrix.	
		\end{flushleft}
		\label{tab:tvd}
	\end{suppTab}
}

Let us now discuss some interesting features of Table \ref{tab:tvd}:

\begin{itemize}
	\item[(i)] If we restrict ourselves to a fixed set of inputs, then for a given number of photons $n$, adding extra modes does not seem to help. In fact, the best TVD in this case for $m>n$ seems to be obtained by embedding an $n \times n$ matrix in the $m \times m$ one that acts as the identity in the remaining modes. As an example, consider the case $n=2$ - the best TVD is the same as obtained by the HOM effect, and the corresponding unitary is simply a beam splitter between two of the modes. 
	\item[(ii)] The same as above naturally does not hold when we average over all inputs. In this case, considering a smaller matrix embedded in the larger one is clearly only optimal for a small fraction of the inputs. A notable example is for the case of 3 photons in 4 modes. In that case, embedding a 3-mode Fourier transform is optimal for the single-input scenario (case $U_5$), but the optimal interferometer for the average case is $U_4$, which consists of a cascade of one 50:50 beam splitter between two modes (say, pair $\{1,2\}$) followed by two parallel beam splitters (between pairs $\{1,3\}$ and $\{2,4\}$).
	\item[(iii)] Comparing only the Fourier ($F_m$) and Sylvester ($S_m$) matrices, for the $n=2$ case, we see that if we only look at a single input they are always tied. However, when looking at the average case, the value for the Sylvester is unchanged, but the Fourier decreases. This is a consequence of the fact that, while the Sylvester matrix is symmetric over all inputs, the Fourier matrix has different TVDs for different sets of inputs, as observed in Fig. 4 in the main text.
	
\end{itemize}

The notable matrices in Table \ref{tab:tvd} are (up to permutation of the modes) as follows:

\begin{itemize}
	\item[$U_1$:] A 50:50 beam splitter between $\{1,2\}$ followed by another between $\{2,3\}$.
	\item[$U_2$:] It is the matrix
	\begin{suppEq}
		U_2 = 
		\begin{pmatrix} 
			1/3 & 2/3 & 2/3 \\
			2/3 & -2/3 & 1/3 \\
			2/3 & 1/3 & -2/3
		\end{pmatrix}
	\end{suppEq}
	\item[$U_3$:] A single 50:50 beam splitter, with identity on the remaining mode.
	\item[$U_4$:] A 50:50 beam splitter between $\{1,2\}$ followed by two parallel ones, between $\{1,3\}$ and $\{2,4\}$.
	\item[$U_5$:] A 3-mode Fourier transform, with identity on the remaining mode.
	\item[$U_6$:] Two parallel 50:50 beam splitters, between $\{1,3\}$ and $\{2,4\}$.
	\item[$U_7$:] A single 50:50 beam splitter, with identity on the remaining modes.
	\item[$U_8$:] A 3-mode Fourier transform, with identity on the remaining modes.
	\item[$U_9$:] A single 50:50 beam splitter, with identity on the remaining modes.
	\item[$U_{10}$:] A 4-mode Fourier transform, with identity on the remaining modes.
\end{itemize}

\section{Tomography of the unitary matrices}
\label{supp_sec2}

Several algorithms have been developed to perform the reconstruction of a unitary linear transformation \cite{Crespi16,Laing12,Lobino2008,Peruzzo11,Rahimi2011,Rahimi2013,Tillmann2016}. One first difference among the various procedures resides in the amount of exploited knowledge about the physical structure of the interferometer. The reconstruction can either be performed considering the device as a black box, whose only known information is the $n$-photon input-output relation \cite{Laing12,Lobino2008,Peruzzo11,Rahimi2011,Rahimi2013}, or considering the device as a complex structure whose elementary constituents' arrangement is partially or completely known \cite{Crespi16,Tillmann2016}. Here, we adopt the second approach for the reconstruction of the Sylvester matrices. With this approach, the retrieval of the unitary transformation is based on the characterization of the constituent optical elements, that is, beam-splitters and phase shifters. 

The advantages of the algorithm are:

\begin{description} 
	\item[Significance] Given two different reconstructed matrices having the same fidelity with the ideal one, a unitary transformation reconstructed by adopting this approach is closer to the implemented one. Indeed, the constraints on the architecture can highlight possible internal symmetries otherwise ignored by other methods.
	\item[Efficiency] Any reconstruction procedure requires the estimate of all the moduli and phases of the unitary transformation implemented. However, by exploiting the a priori knowledge on the inner structure, the number of necessary measurements can be reduced to the minimum number of input-output relations sufficient to retrieve all the parameters of the involved optical elements.
	\item[Unitarity] Other algorithms which treat the system as a black box may in general give a non-unitary matrix as output of the reconstruction \cite{Laing12}. The subsequent need for recovering a unitary transformation may thus alter, to some extent, the result of the reconstruction process.
	\item[Characterization] Ultimately, further information is obtained by focusing on the properties of the elementary optical components. Specifically, this reconstruction method permits to investigate the tolerance to their unavoidable fabrication imperfections, by comparing their retrieved characteristics with those of the ideal ones.
\end{description} 

To perform the reconstruction, i.e. to characterize the single components within the integrated architecture, it is necessary to gain information on both the transmittivities $\{\tau\}$ and the relative phases $\{\phi\}$ associated to each element. Indeed, it can be shown that the element $[U_{\mathrm{S}}^{(2^n)}]_{i,j}$ of a $d$-dimensional Sylvester transformation ($d=2^{n}$), implemented using the Barak and Ben-Aryeh decomposition \cite{Barak07}, can be parametrized in $(\tau,\phi)$ as

\begin{suppEq}
	\label{eq:decomposition}
	[U_{\mathrm{S}}^{(2^n)}]_{i,j} = \sum_{k_{1}=1}^{2^n} ..\sum_{k_{n-1}=1}^{2^n} \: \prod_{s=0}^{n-1} \;{\textbf{L}_{k_s,k_{s+1}}^{(n-s)} },
\end{suppEq}
where the matrices $\textbf{L}_{k_s,k_{s+1}}^{(t)}$ describe the action of the $n=\log{d}$ layers, each consisting, in turn, of a layer of $ d/2$ beam-splitters $\textbf{B}_{k_s,k_{s+1}}^{(t)}$, placed between the modes $(k_s, k_{s+1})$, and $ d/2$ phase shifters $e^{\imath \phi_m^{(t)}}$  placed on the higher-index mode of the same pairs $(m=\max[k_s,k_{s+1}])$

\begin{suppEq}
	\label{eq:not1}
	\textbf{L}_{k_s,k_{s+1}}^{(t)}= \textbf{B}_{k_s,k_{s+1}}^{(t)}\: e^{\imath \phi_{\max[k_s,k_{s+1}]}^{(t)}},
\end{suppEq}
having also absorbed the indexes $\textit{i}$ and $\textit{j}$ in $\textit{k}_0$ and $\textit{k}_n$ respectively. Here, the matrix associated to the layer of beam-splitters has the form

\begin{suppEq}
	\textbf{B}_{k_s,k_{s+1}}^{(n-s)}\equiv
	\left\{
	\begin{array}{ll}
		\ \tau_{s,k_{s}}^{(n)} \qquad \qquad \quad \:\:\, \: k_s=k_{s+1} \\
		\ \imath \, \sqrt{1-\tau_{s,k_{s}}^{(n)\;2}}  \qquad (k_s,k_{s+1}) \in \big\{(\alpha, \beta)\big\}^{(n-s)} \\
		\ \\
		\ 0  \hspace{7.1em} \textrm{otherwise}\\
	\end{array}
	\right.            
	\label{eq:fourier}                        
\end{suppEq}
where, for a given layer $s$ in a $2^n$-dimensional interferometer, $\{(\alpha, \beta) \}^{(n-s)}$ is the set of pairs of modes interacting at each step of the fast implementation:
\begin{suppEq}
	\Big\{(\alpha, \beta) \Big\}^{(t)}=\Big\{ \left(\: a+2^{t}\: b,\: a+2^{t}\: b+2^{t-1}\: \right) \Big\}
\end{suppEq}
with $a \in {\{1,..,2^{n-s+1} \}}\,,\; b \in { \{ 1,..,2^s\}}$.

\begin{figure*}[htp]
	\centering
	\includegraphics[width=0.9\textwidth]{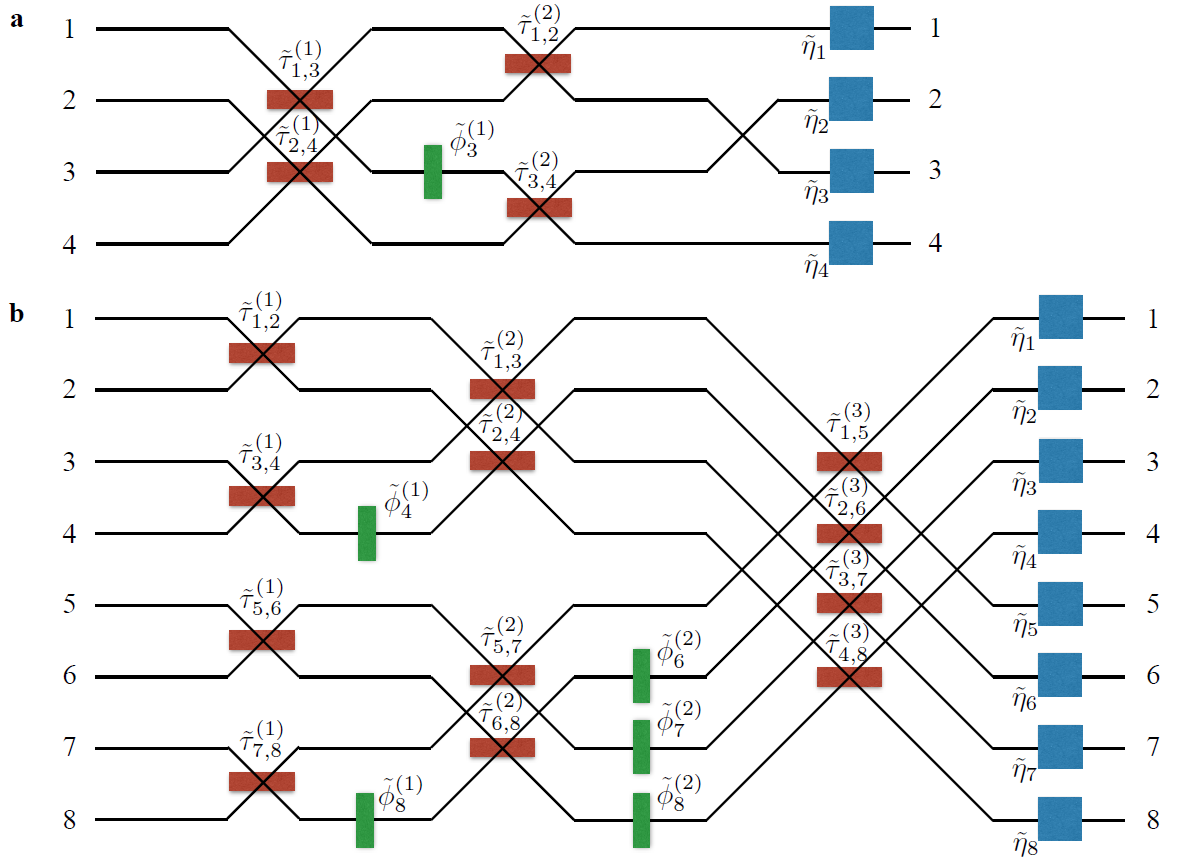}
	\caption{Internal structure of the 4-mode device (\textbf{a}) and the 8-mode device (\textbf{b}). A different notation for $\{\tau\}$ and $\{\phi\}$ is adopted with respect to Eqs. (\ref{eq:not1}) and (\ref{eq:fourier}). Red boxes: beam-splitters. Green boxes: fabrication phases relevant for the algorithm. Blue boxes: relative output losses. For the Sylvester transformations, all beam-splitters have transmittivities $\tau_{\mathrm{id}} =2^{-1/2}$ and all phases are 0.}
	\label{fig:schemechips}
\end{figure*}

The reconstrution algorithm is made up of two independent stages, corresponding to the separate retrieval of the transmittivities $\{\tau\}$ and the phases $\{\phi\}$. The estimate of the two sets of parameters is performed via a maximum-likelihood method. In Figure \ref{fig:schemechips} we report the schemes of the internal structure of the implemented interferometers, considered as a priori knowledge for the algorithm.

\subsection*{Estimate of the transmittivities}
The dependence of the moduli on the transmittivities $\{ \tau\}$ is outlined in ($\ref{eq:fourier}$). Information on the $ \vert U_{\mathrm{S}}^{(d)}\vert_{i,j}$ can be obtained directly with single-photon measurements, recalling that the scattering probability from modes $i$ to $j$ is indeed exactly $\vert U_{\mathrm{S}}^{(d)}\vert_{i,j}^2$. Thus, by experimentally measuring all the $d^{2}$ scattering probabilities $[\tilde{P}_{\mathrm{S}}^{(d)}]_{i, j}$, it is possible to look for the set  $ \{ \tau\}$ that best reproduces the values registered. This can be done, via a maximum likelihood procedure, by numerically minimizing the $\chi_\tau^2$ quantity

\begin{suppEq}
	\chi_\tau^2={\sum_{i,j=1}^{d} \left( \frac{[\tilde{P}_{\mathrm{S}}^{(d)}]_{i, j}-\vert U_{\mathrm{S}}^{(d)}\vert_{i,j}^2}{[\sigma_{\tilde{P}_{\mathrm{S}}}^{(d)}]_{i,j}}\right)^2},
\end{suppEq}
where $\sigma_{\tilde{P}_{\mathrm{S}}}^{(d)}$ are the experimental errors associated to the measured probabilities $\tilde{P}_{\mathrm{S}}^{(d)}$. However, the set $ \{ \tau\}$  found in this way is insensitive to possible external losses in the output of the device (see Figure \ref{fig:schemechips}). Such unavoidable relative losses modify the transformation actually implemented, entailing biases in the distribution of the moduli. To overcome this issue it is sufficient to introduce a second, non-unitary diagonal transformation $D(\eta_1,\eta_2,...,\eta_{d})$ acting after the device. Thus, defining

\begin{suppEq}
	M_{\mathrm{S}}^{(d)}=D(\eta_1,\eta_2,...,\eta_{d}) U_{\mathrm{S}}^{(d)} = 
	\begin{pmatrix} 
		\eta_{1} & 0 & \ldots & 0 \\
		0 & \eta_{2} & \ldots & 0 \\
		\vdots & \vdots & \ddots & \vdots \\
		0 & 0 & \ldots & \eta_{d}
	\end{pmatrix} U_{\mathrm{S}}^{(d)},
\end{suppEq}
and including the $\{ \eta\}$ in the parametrization of the unitary, it is possible to obtain an estimate of the actual relative losses $\{\tilde{\eta}\}$ involved, as well as a set of transmittivities $ \{ \tilde{\tau}\}$ which properly takes them into account.

The retrieved values of the output losses for the 4-mode and the 8-mode devices, as well as the transmittivities $ \{ \tilde{\tau}\}$, are reported in Table \ref{tab:parameters1}. Note that since the method permits reconstruction of relative losses only, one of the parameters must be taken as a reference (for instance, $\eta_{1}$).

\begin{suppTab}[ht!]
	\centering
	\resizebox{\columnwidth}{!}{
	\begin{tabular}{|c|c||c|c|}
		\hline 
		4-mode device: $\{ \tilde{\eta} \}$ & 4-mode device: $\{ \tilde{\tau} \}$ & 8-mode device $\{ \tilde{\eta} \}$ & 8-mode device $\{ \tilde{\tau} \}$ \\
		\hline
		$\tilde{\eta}_{1} = 1$ & $\tilde{\tau}^{(1)}_{1,3} = 0.6879 \pm 0.0003$ & $\tilde{\eta}_{1} = 1$ & $\tilde{\tau}^{(1)}_{1,2} = 0.669 \pm 0.008$\\
		$\tilde{\eta}_{2} = 1.134 \pm 0.002 $ & $\tilde{\tau}^{(1)}_{2,4} = 0.7195 \pm 0.0003$ & $\tilde{\eta}_{2} = 0.94 \pm 0.02$ & $\tilde{\tau}^{(1)}_{3,4} = 0.621 \pm  0.007$\\
		$\tilde{\eta}_{3} = 1.147 \pm 0.002$ & $\tilde{\tau}^{(2)}_{1,2} = 0.7139 \pm 0.0003$ & $\tilde{\eta}_{3} = 1.07 \pm 0.02$ & $\tilde{\tau}^{(1)}_{5,6} = 0.639 \pm  0.008$\\
		$\tilde{\eta}_{4} = 0.961 \pm 0.002$ & $\tilde{\tau}^{(2)}_{3,4} = 0.7031 \pm 0.0003$ & $\tilde{\eta}_{4} = 0.93 \pm 0.02$ &  $\tilde{\tau}^{(1)}_{7,8} = 0.630 \pm  0.008$\\
		& & $\tilde{\eta}_{5} = 0.96 \pm 0.02$ & $\tilde{\tau}^{(2)}_{1,3} = 0.750 \;\pm  0.007$\\
		& & $\tilde{\eta}_{6} = 1.05 \pm 0.02$ & $\tilde{\tau}^{(2)}_{2,4} = 0.715 \;\pm  0.007$\\
		& & $\tilde{\eta}_{7} = 1.01 \pm 0.02$ & $\tilde{\tau}^{(2)}_{5,7} = 0.755 \;\pm  0.007$\\
		& & $\tilde{\eta}_{8} = 1.02 \pm 0.02$ & $\tilde{\tau}^{(2)}_{6,8} = 0.729 \;\pm  0.006$\\
		& & & $\tilde{\tau}^{(3)}_{1,5} = 0.748 \;\pm  0.008$\\
		& & & $\tilde{\tau}^{(3)}_{2,6} = 0.723 \;\pm  0.007$\\
		& & & $\tilde{\tau}^{(3)}_{3,7} = 0.774 \;\pm  0.006$\\
		& & & $\tilde{\tau}^{(3)}_{4,8} = 0.755 \;\pm  0.007$\\
		\hline
	\end{tabular}
}
	\caption{Relative losses $\{ \tilde{\eta} \}$ and transmittivities $\{ \tilde{\tau} \}$ reconstructed from the first step of the algorithm for the 4- and 8-mode devices.}
	\label{tab:parameters1}
\end{suppTab}

The values of $ \{ \tilde{\tau}\}$ are to be compared with the expected one $(\tau_{\mathrm{id}}=2^{-1/2})$. In all estimates, the experimental errors are obtained by a MonteCarlo approach on the reconstruction process.

Let us now define $U_{\tau}^{(d)}$ as the temporary matrix obtained by inserting the $\{ \tilde{\tau}\}$ in the decomposition (\ref{eq:decomposition}) of $U_{S}^{(d)}$.

\subsection*{Estimate of the phases $ \psi_{i,j}$ }

Once we have the estimate of the set $\{ \tilde{\tau}\}$, it is possible to apply the same procedure to find the values of the phases $\{ \tilde{\phi} \}$ that best reproduce further experimental measurements. This is done by minimizing a second $\chi_\phi^2$ quantity, which compares the measured visibilities ($\tilde{V}_{i, j}^{m,n}$) of the two-photon HOM dips/peaks with the ones ($V_{i, j}^{m,n}$) expected from the unitary $U_{\tau}^{(d)}$. Note that the expected values are parametric in the unknown phases $\{ \tilde{\phi} \}$. The visibilities are defined as $V_{i,j}^{m,n}=(D_{i,j}^{m,n}-Q_{i,j}^{m,n})/D_{i,j}^{m,n}$, being $D_{i,j}^{m,n}$ the transition probability from inputs $(i,j)$ to outputs $(m,n)$ with distinguishable particles and $Q_{i,j}^{m,n}$ the one with indistinguishable photons. The visibilities $V_{i,j}^{m,n}$ are measured by recording the output coincidences as a function of the relative delay $\Delta \tau$, and by fitting the experimental interference pattern with the function $C_{i,j}^{m,n}(\Delta \tau) = B_{i,j}^{m,n} [1 - V_{i,j}^{m,n} \exp(-\Delta \tau^{2}/(2 \sigma_{\tau}^{2})]$. The $\chi_\phi^2$ quantity to be minimized for the reconstruction algorithm is defined as 

\begin{suppEq}
	\chi_\phi^2={\sum_{i,j,m,n}^{d} \left( \frac{\tilde{V}_{i, j}^{m,n}-V_{i,j}^{m,n}}{\sigma_{\tilde{V}_{i,j}^{m,n}}}\right)^2},
\end{suppEq}
where the sum is extended over all the measured input-output combinations.
Note that visibilities are independent of the output losses. Here, for each visibility, the error $\sigma_{\tilde{V}_{i,j}^{m,n}}$ is obtained from the fitting procedure. Conveniently choosing the two-photon input states to inject, whose visibilities depend on all the relevant phases, it was in principle possible to complete the reconstruction using only one state. However, for strengthening the reconstruction, for both devices we measured an overcomplete set of data. In the 4-mode case, we recorded all possible 36 collision-free two-photon visibilities. In the 8-mode case, all collision-free visibilities for 8 two-photon input states $ \{ (2,5),(2,6),(2,8),(3,7),(3,8),(5,7),(5,8),(6,8) \}$ have been recorded, for a total number of 224 visibilities.

The estimated values for the phases are reported in Table \ref{tab:parameters2}.
\begin{suppTab}[htbp]
	\centering
	\resizebox{\columnwidth}{!}{
	\begin{tabular}{|c||c|}
		\hline 
		4-mode device: $\{ \tilde{\phi} \}$ & 8-mode device $\{ \tilde{\phi} \}$\\
		\hline
		$\phi^{(1)}_{3} = (+0.229 \pm 0.004)$ rad & $\phi^{(1)}_{4} = (+0.38 \pm 0.02)$ rad\\
		& $\phi^{(1)}_{8} = (+0.36 \pm 0.01)$ rad\\
		& $\phi^{(2)}_{6} = (-0.19 \pm 0.03)$ rad\\
		& $\phi^{(2)}_{7} = (+0.43 \pm 0.02)$ rad\\
		& $\phi^{(2)}_{8} = (-0.04 \pm 0.03)$ rad\\
		\hline
	\end{tabular}
}
	\caption{Relevant phases $\{\tilde{\phi} \}$ reconstructed from the second step of the algorithm for the 4- and 8-mode devices.}
	\label{tab:parameters2}
\end{suppTab}

Note that, thanks to the internal symmetries in the evolution through the interferometer, it is possible to describe the action of the fast Sylvester matrix with fewer phases than those actually present in the devices (see scheme of Figure \ref{fig:schemechips}). First, input phases are irrelevant when the system is injected with Fock states, and thus can be all set to 0. In the case of the 4-mode device, for instance, only one phase is relevant for the algorithm, and we can thus set $\phi_{4}^{(1)}=0$ for the sake of simplicity. In the case of the 8-dimensional Sylvester transformation, the functional dependence on all the 4+4 phases of the two internal layers can be compressed in that on only 5 phases $(\phi^{(1)}_{3} = \phi^{(1)}_{7}=\phi^{(2)}_{5}= 0)$, eventually involving differences between them.

Once all the parameters $\{\tilde{\tau}\}$ and $\{\tilde{\phi}\}$ are known, the actual implemented unitary transformations $\tilde{U}_{\mathrm{fS}}^{(d)}$ can be obtained by inserting the obtained values in the decomposition (\ref{eq:decomposition}).

\subsection*{Results}

\begin{figure*}[htp]
	\centering
	\includegraphics[width=0.9\textwidth]{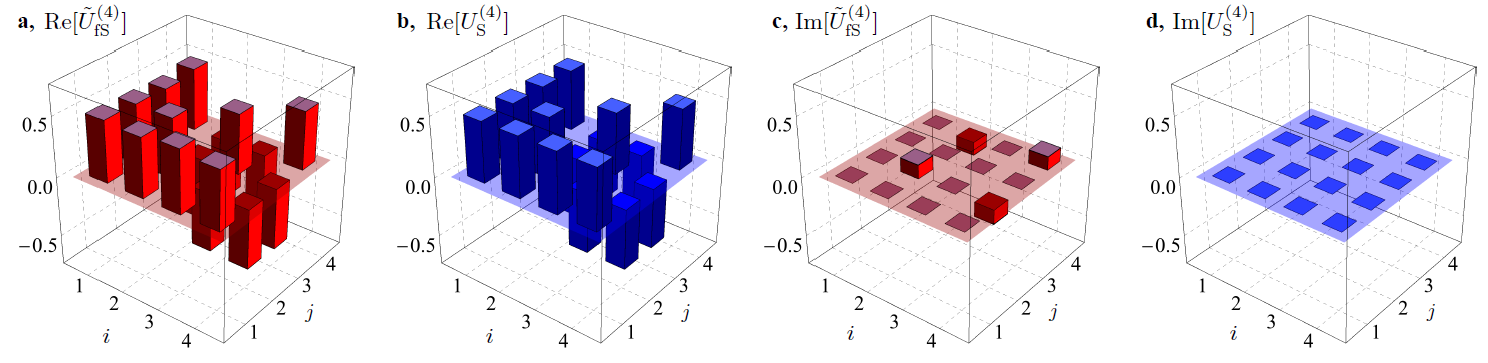}
	\caption{Results of the unitary reconstruction process for the 4-mode device. \textbf{a} Real part and \textbf{c} imaginary part of the implemented interferometer $\tilde{U}_{\mathrm{fS}}^{(4)}$. Lighter regions in the bars represent the error in the reconstruction process. \textbf{b} Real part and \textbf{d} imaginary part of the Sylvester matrix $U_{\mathrm{S}}^{(4)}$.}
	\label{fig:results_tomo4}
\end{figure*}

\begin{figure*}[htp]
	\centering
	\includegraphics[width=0.9\textwidth]{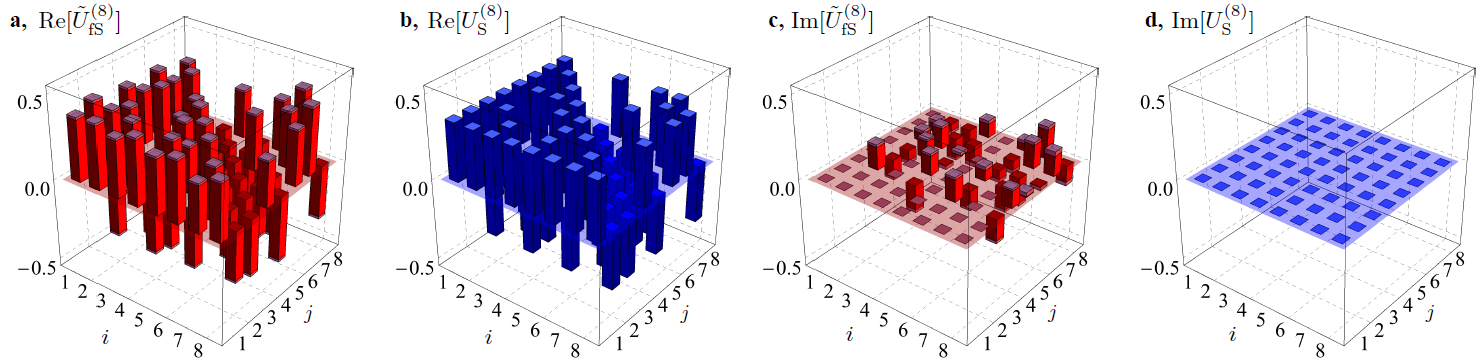}
	\caption{Results of the unitary reconstruction process for the 8-mode device. \textbf{a} Real part and \textbf{c} imaginary part of the implemented interferometer $\tilde{U}_{\mathrm{fS}}^{(8)}$. Lighter regions in the bars represent the error in the reconstruction process. \textbf{b} Real part and \textbf{d} imaginary part of the Sylvester matrix $U_{\mathrm{S}}^{(8)}$.}
	\label{fig:results_tomo8}
\end{figure*}

The real and imaginary parts of the reconstructed transformation are compared with the Sylvester transformations in Figure \ref{fig:results_tomo4} (for the 4-mode device) and in Figure \ref{fig:results_tomo8} (for the 8-mode device). The fidelities $F^{(d)}=(1/d) \vert \mathrm{Tr}(U_{\mathrm{S}}^{(d) \, \dag} \, \tilde{U}_{\mathrm{fS}}^{(d)})\vert$ with the ideal 4- and 8-dimensional Sylvester transformation are respectively $F^{(4)} =0.99807 \pm 0.00005$ and $F^{(8)}=0.9813 \pm 0.0005$, showing the good quality of the fabrication process.

\section{Total variation distance for 3D interferometers with fast architecture}
\label{supp_sec3}

Here we discuss the TVD between the distributions with indistinguishable and distinguishable particles when the interferometer is implemented with fast architecture. The layout for the 4-mode and 8-mode devices are shown in Figure \ref{fig:schemechips}. We restrict our attention to the case where all directional couplers are symmetric (and thus with transmittivities $\tau_{s,k_{s}}^{(n)}=2^{-1/2}$). Sylvester or Fourier interferometers can be recovered by inserting appropriate values for the fabrication phases $\phi_{k}^{(n)}$. 

\begin{figure*}[htp]
	\centering
	\includegraphics[width=0.9\textwidth]{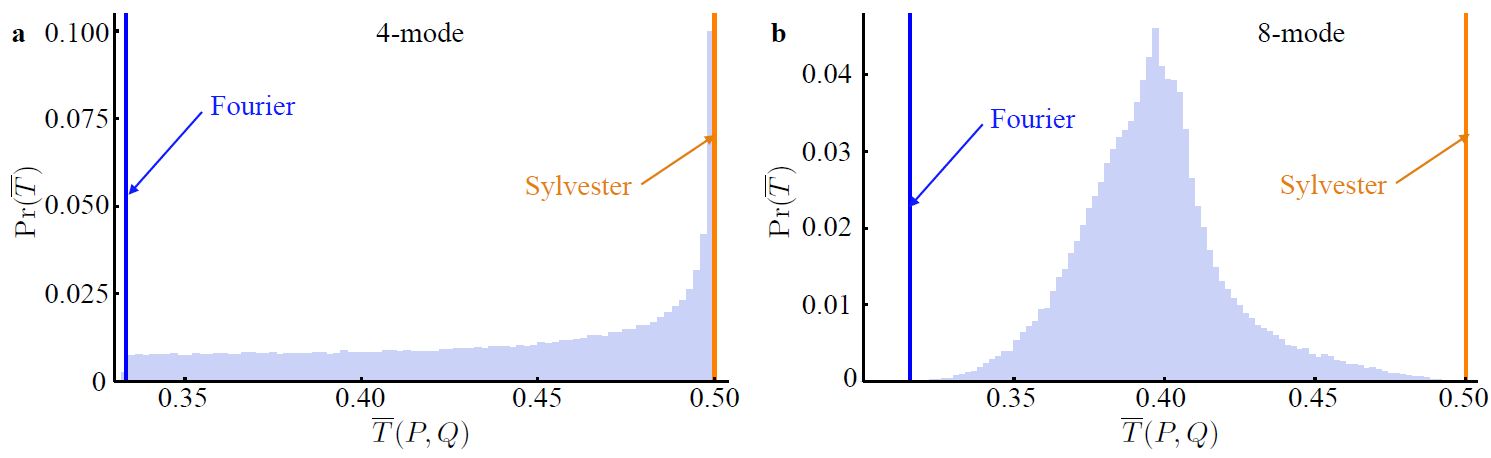}
	\caption{Total variation distance in the multiple input configuration for 3D interferometers with fast architecture, with transmittivities $\tau_{s,k_{s}}^{(n)}=2^{-1/2}$ and random, uniformly distributed, phases $\phi_{k}^{(n)}$. Histograms of the TVD for $N=10^5$ unitaries for {\bf a} $n=2$, $m=4$ and {\bf b} $n=2$, $m=8$. Orange vertical bar: Sylvester interferometer. Blue vertical bar: Fourier interferometer.}
	\label{fig:noise}
\end{figure*}

We performed a numerical simulation of the TVD by generating $N=10^{5}$ different unitary transformations according to the Barak and Ben Aryeh architecture, with transmittivities $\tau_{s,k_{s}}^{(n)}=2^{-1/2}$ and random, uniformly distributed, phases $\phi_{k}^{(n)}$ for $n=2$ photons and $m=4,8$ modes. This permits to evaluate the effects of fabrication noise in the output distributions of the implemented interferometers. In the fixed input case, we observe that cyclic inputs always lead to the optimal value $T(P,Q)=0.5$, independently of the values of phases $\phi_{k}^{(n)}$. Fabrication errors in the optical phases within the interferometers will thus not affect the obtained results in this scenario. In the multiple input configuration, on the other hand, we observe a different scenario. In the 4-mode case, the maximum and the minimum values of the TVD are obtained respectively for the Sylvester and the Fourier transformation (see Figure \ref{fig:noise}a). In the 8-mode case, the maximum is still obtained for the Sylvester interferometer, while a (very) small subset of unitaries presents a lower value of the TVD with respect the Fourier transformation (see Figure \ref{fig:noise}b). For the $N=10^{5}$ set of randomly generated unitaries, the obtained minimum is $\overline{T}(P,Q) \sim 0.298$, while the value for the Fourier is $\overline{T}(P,Q) \sim 0.315$.

Our simulations thus indicate that fabrication noise in the implementation of Fourier interferometers will tend to result in an increased value of the TVD, with respect to the ideal Fourier transformation. As we have seen in the main text, this was indeed what was observed experimentally in the two devices we characterized.

\section{Bayesian validation test}
\label{supp_sec4}

The aim is to discriminate between two hypotheses $Q$, corresponding to a Boson Sampling device and $P$, corresponding to the evolution obtained with distinguishable particles. To this end, it is possible to employ a Bayesian test as shown in {(\it 10)}. At the initial stage, no a priori knowledge is assumed on the system. This translates into uniform prior probabilities $\mathrm{Pr}(Q)=\mathrm{Pr}(P)=0.5$. After the measurement of $N_{\mathrm{events}}$ events, the ratio between the conditional probabilities associated to the two hypotheses is updated according to Bayes' rule:
\begin{suppEq}
	R=\frac{\mathrm{Pr}(Q \vert N_{\mathrm{events}})}{\mathrm{Pr}(P \vert N_{\mathrm{events}})} = \prod_{i=1}^{N_{\mathrm{events}}} \left( \frac{q_{i}}{p_{i}} \right),
\end{suppEq}
where $i$ is the index of each sampled event, $q_{x}$ and $p_{x}$ are the corresponding probabilities associated to the two hypotheses $Q$ and $P$ respectively. Thus, the confidence probability for hypothesis $Q$ of indistinguishable particles after $N_{\mathrm{events}}$ reads:
\begin{suppEq}
	\mathrm{Pr}(Q \vert N_{\mathrm{events}}) = \frac{1}{\mathcal{N}} \prod_{i=1}^{N_{\mathrm{events}}} \left( \frac{q_{i}}{p_{i}} \right),
\end{suppEq}
where $\mathcal{N}$ is a normalization constant.

This test can be applied in two different scenarios, namely when the incoming data are generated by a Boson Sampling device (ind) or by distinguishable particles (dis). When the incoming data are generated by indistinguishable particles, $\mathrm{Pr}_{\mathrm{ind}}(Q \vert N_{\mathrm{events}})$ represents the conditional probability associated to hypothesis $Q$ (that is, the correct guess), while $\mathrm{Pr}_{\mathrm{ind}}(P \vert N_{\mathrm{events}})$ represents the probability associated to hypothesis $P$ (that is, the wrong guess). Analogous definitions are obtained when the data are generated by distinguishable particles, where $\mathrm{Pr}_{\mathrm{dis}}(Q \vert N_{\mathrm{events}})$ corresponds to the wrong guess and $\mathrm{Pr}_{\mathrm{dis}}(P \vert N_{\mathrm{events}})$ to the correct guess.

The confidence probability of the test can be then defined as:
\begin{suppEq} 
	P_{\mathrm{conf}} = \frac{1}{2} \left[\mathrm{Pr}_{\mathrm{ind}}(Q \vert N_{\mathrm{events}}) + \mathrm{Pr}_{\mathrm{dis}}(P \vert N_{\mathrm{events}})\right].
\end{suppEq} 
Conversely, $P_{\mathrm{err}} = 1 - P_{\mathrm{conf}}$ represents the average error probability of the test.

\section{Bayesian test on the experimental data}
\label{supp_sec5}

Here we provide more details on the application of the Bayesian validation test on the experimental data. For each input state, $M_{\mathrm{sample}} = 10^{4}$ distinct samples were generated numerically from the experimentally measured two-photon distributions with indistinguishable photons and distinguishable particles, the latter obtained by applying a suitable time delay $\Delta \tau$ between the photons larger than their coherence time. For each data sample, Gaussian noise was introduced in the distribution according to the experimental errors. Then, for each data sample the Bayesian test was applied using event data sets of increasing size $N_{\mathrm{events}}$. In the 4-mode case, the full set of input states have been measured in the no-collision subspace, and the cumulative contribution of collision events has been included in an extra bin (see Section ``Measurement of the TVD'' of main text). The confidence probability $P_{\mathrm{conf}}$ is thus averaged over all input configurations. In the 8-mode case, three input states have been measured by detecting also the collision events. For the Sylvester interferometer, all input states are equivalent, and thus an average with uniform weights has been performed. 

\section*{Partial photon distinguishability in the binary Bayesian test}
\label{supp_sec6}

\begin{figure*}[htp!]
	\centering
	\includegraphics[width=0.8\textwidth]{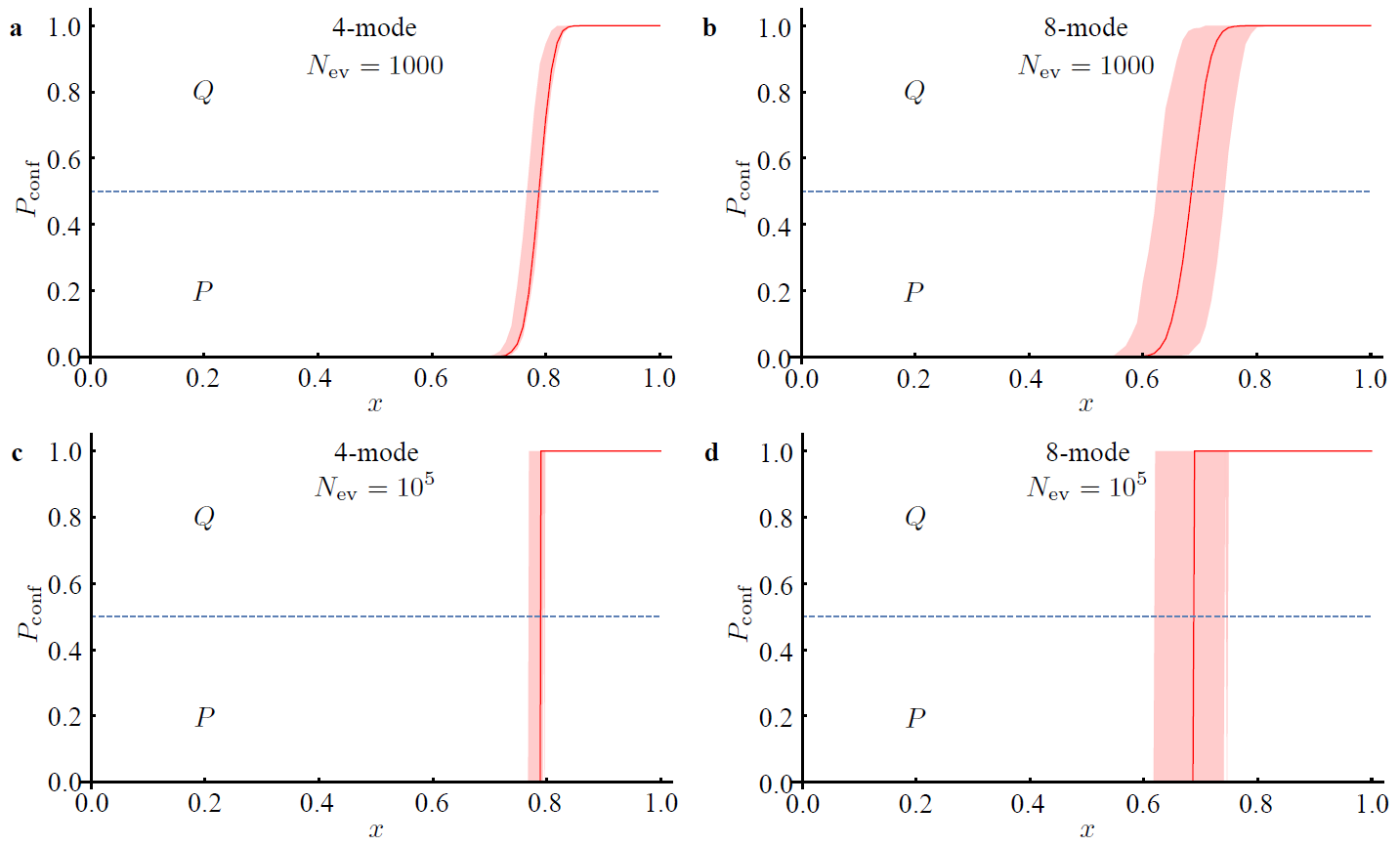}
	\caption{Confidence probability $P_{\mathrm{conf}}$ as a function of the photon distinguishability $x$, averaged over $10^4$ distinct samples generated from the distribution calculated from the reconstructed unitary transformations $\tilde{U}_{\mathrm{fS}}^{(4)}$ and $\tilde{U}_{\mathrm{fS}}^{(8)}$. After $N_{\mathrm{ev}} \sim 1000$ events: {\bf a}, $4$-mode Sylvester interferometer and {\bf b}, $8$-mode Sylvester interferometer. After $N_{\mathrm{ev}} \sim 10^5$ events: {\bf c}, $4$-mode Sylvester interferometer and {\bf d}, $8$-mode Sylvester interferometer. Shaded regions: fixed input configuration, interval comprising the curves for each individual state. Solid Lines: multiple input configuration.}
	\label{fig:partial}
\end{figure*}

Here we discuss the role of partial photon distinguishability in the Bayesian test when discriminating between hypotheses $P$ and $Q$. More specifically, we analyzed the case when the experimental data are generated from photon sources with partial photon distinguishability. In the two-photon case, this means that the actual two-photon distribution can be approximated as a convex combination $h_{i}(x) = x q_{i} + (1-x) p_{i}$, where $x \in [0,1]$ is a real parameter quantifying the degree of indistinguishability between the photons ($x=1$ corresponds to the perfect indistinguishable case). We keep the $Q$ ($P$) hypotheses to be the distributions with perfectly indistinguishable (distinguishable) photons, while we allow partial distinguishability on the data to determine the threshold on the indistinguishability $x$ below which the Bayesian test inverts its outcome. To this end, we performed some numerical simulations shown in Figure \ref{fig:partial} for the $4$-mode and $8$-mode implemented Sylvester interferometers by applying the test to numerically generated data samples calculated from the reconstructed unitaries $\tilde{U}_{\mathrm{fS}}^{(4)}$ and $\tilde{U}_{\mathrm{fS}}^{(8)}$. We then evaluated the asymptotic confidence probability $P_{\mathrm{conf}}$ after collecting a large number of events ($N_{\mathrm{ev}} \sim 1000$), as a function of the parameter $x$. When $P_{\mathrm{conf}}>0.5$ ($P_{\mathrm{conf}}<0.5$), the full experiment is attributed to hypothesis $Q$ ($P$). We observe that, in the $4$-mode case, data with $x > 0.788$ are still assigned to indistinguishable particles even in presence of partial distinguishability ($x > 0.685$ in the $8$-mode case).

\section*{Bayesian test on scattershot Boson Sampling experimental data}

In this section we describe the Bayesian inference method applied in the main text on the scattershot Boson Sampling data, and describe a related alternative approach based on binary likelihood ratio tests.

\subsection*{Bayesian inference to ascertain photon indistinguishability}

Bayesian inference has been applied in the main text on the scattershot Boson Sampling data. More specifically, the binary set of hypotheses $Q$ and $P$ is replaced by the convex combination $h_{i}(x) = x q_{i} + (1-x) p_{i}$ described in section \ref{supp_sec6}. Thus, experimental data are assumed to be described by the combination $h_{i}$ for a given value $x_{\mathrm{true}}$ of the indistinguishability parameter $x$. Then, the value of $x$ for the collected data is retrieved from a given data sample by performing single-parameter Bayesian estimation. Namely, a uniform prior $\mathcal{P}(x)$ is assumed for $x$, quantifying the total a-priori ignorance on the parameter. The conditional probability of $x$ after $N$ measured events is obtained from the Bayes rule:
\begin{suppEq}
	\mathcal{P}(x \vert i_{1}, \ldots, i_{N}) = \frac{\mathcal{P}(i_{1}, \ldots, i_{N} \vert x) \mathcal{P}(x)}{\mathcal{N}},
\end{suppEq}
where $(i_{1}, \ldots, i_{N})$ is the recorded data sample, $\mathcal{P}(i_{1}, \ldots, i_{N} \vert x)$ is the conditional probability of obtaining the data sample for a given value of $x$, and $\mathcal{N}$ is a normalization constant. The estimated value of the parameter (and its associated error) after $N$ measured events are retrieved from the distribution $\mathcal{P}(x \vert i_{1}, \ldots, i_{N})$ as:
\begin{suppEq}
	x_{\mathrm{est}} = \int_{0}^{1} x \mathcal{P}(x \vert i_{1}, \ldots, i_{N}) dx,
\end{suppEq}
and
\begin{suppEq}
	\sigma_{\mathrm{est}} = \left[\int_{0}^{1} (x-x_{\mathrm{est}})^{2} \mathcal{P}(x \vert i_{1}, \ldots, i_{N}) dx\right]^{1/2}.
\end{suppEq}

The method has been applied on scattershot Boson Sampling experimental data, and the results are shown in Fig. 6b-c of the main text. More specifically, starting from the overall collected data sample of $\sim 17000$ events, we generated numerically $100$ different data sequences, in order to evaluate the variability of the test when the data sequence is changed. By using the overall data set, the final estimated value of $x$ is $\tilde{x}_{\mathrm{est}} = 0.738 \pm 0.004$, which is compatible with the one retrieved from Hong-Ou-Mandel interference with two of the PDC sources in a 50/50 beam-splitter: $x_{\mathrm{HOM}} = 0.79 \pm 0.06$.

\subsection*{Convex hypothesis test with likelihood ratios}

A different approach based on binary likelihood ratio tests can be employed to identify the value of $x$ that better describes the experimental data. As in the previous case, the hypothesis is that data are described by the convex combination $h_{i}(x) = x q_{i} + (1-x) p_{i}$ for some value $x_{\mathrm{true}}$ of $x$. This approach relies on the likelihood ratio test described in section \ref{supp_sec4}, and is divided in two stages.

\textbf{Stage A}. The first stage of the test exploits experimental data to identify a threshold value $x_{\mathrm{th}}$ where the binary decision test becomes ambiguous. More specifically, let us consider the case where the experimental data pass the binary test for the hypothesis $Q$. Then, the test is performed with the same data by fixing the hypothesis $Q$ and by using as alternative hypothesis $H(x)$. The test is repeated by tuning the value of $x$ until the ratio $R(x)$ between the hypotheses $Q$ and $H(x)$ lead to an ambiguous results:
\begin{suppEq}
	R(x)=\prod_{i=1}^{N_{\mathrm{events}}} \frac{q_{i}}{x q_{i} + (1-x) p_{i}}.
\end{suppEq}
The value of $x_{\mathrm{th}}$ then corresponds $R(x_{\mathrm{th}}) = 1$. Analogously, if experimental data pass the test for the hypothesis $P$ the test is performed between $P$ and $H(x)$ thus evaluating the ratio:
\begin{suppEq}
	R(x)=\prod_{i=1}^{N_{\mathrm{events}}} \frac{x q_{i}+(1-x) p_{i}}{p_{i}}.
\end{suppEq}

\textbf{Stage B}. The second stage employs simulated data from the unitary transformation. The test is now performed by keeping fixed the two hypothesis to $Q$ and $H(x_{\mathrm{th}})$ (or $P$ and $H(x_{\mathrm{th}})$ if the data pass the initial test for $P$). Numerically simulated data sample of size $N_{\mathrm{sim}}$ are generated from the distribution $H(y)$ for different value of $y$, and the binary likelihood ratio test is performed by evaluating:
\begin{suppEq}
	R'= \prod_{i=1}^{N_{\mathrm{sim}}} \frac{q_{i}}{x_{\mathrm{th}} q_{i} + (1-x_{\mathrm{th}}) p_{i}} \; 
\end{suppEq}

\noindent or

\begin{suppEq}
 \; R'= \prod_{i=1}^{N_{\mathrm{sim}}} \frac{x_{\mathrm{th}} q_{i} + (1-x_{\mathrm{th}}) p_{i}}{p_{i}}
\end{suppEq}
This procedure is repeated by tuning $y$ until a value $y'$ is obtained for which the test is ambiguous ($R'=1$). This value of $y'$ is an estimate of the parameter $y$. Intuitively, this second stage determines the value of the photon indistinguishability that leads to an ambiguous result for the same value of $x_{\mathrm{th}}$ obtained from experimental data.

\begin{figure*}[htbp!]
	\centering
	\includegraphics[width=1\textwidth]{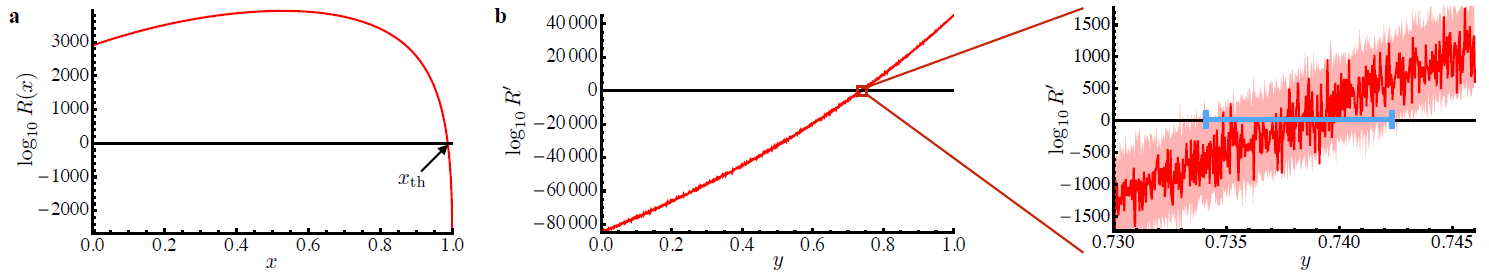}
	\caption{Evaluation of the indistinguishability $x$ through the convex hypothesis test based on the likelihood ratio approach. {\bf a}, First stage of the test with the experimental data to determine the value of $x_{\mathrm{th}}$. {\bf b}, Second stage of the test with numerically simulated data. Left figure: plot of $\log_{10}R'$ as a function of $y$ (indistinguishability of the numerically generated) data for the full interval $y \in[0;1]$. Right figure: highlight in an interval close to $R'=1$. Solid line: variability of $R'$ as a function of $y$ when each point is evaluated from a single-data set. Shaded region: interval for $R'$ obtained by generating $100$ distinct data samples for each value of $y$. The interval included within the blue ruler identifies the interval for the estimate $y'$.}
	\label{fig:convex}
\end{figure*}

We have then applied the test to the experimentally measured scattershot Boson Sampling data. In Stage A, we obtained $x_{\mathrm{th}} \simeq 0.9872$ by using the full data sample (see Figure \ref{fig:convex}a). In Stage B, we performed the likelihood test between hypotheses $P$ and $H(x_{\mathrm{th}})$ by generating samples of $N_{\mathrm{sim}} = 10^{5}$ data from the distributions $H(y)$ for different values of $y$. The distributions $p_{i}$ and $q_{i}$ for the convex combination are calculated from the reconstructed unitary transformation $\tilde{U}_{\mathrm{fS}}^{(4)}$. To give an estimate of the error associated to $y'$ and take into account the effect of using a finite number of samples, in Stage B we generated $100$ distinct data samples of size $N_{\mathrm{sim}}$ for each value of $y$. This allows to obtain an interval for the estimate $y'$. By naming $y'=x_{\mathrm{LR}}$, we obtain $x_{\mathrm{LR}} \in [0.734;0.742]$ (see  Figure \ref{fig:convex}b). This interval is compatible with the value $\tilde{x}_{\mathrm{est}}$ obtained via Bayesian inference.

\section{Experimental tools}

\subsection*{Fabrication of integrated optics devices} 

Waveguide devices are inscribed by femtosecond laser writing technology in alumino-borosilicate glass substrates (EAGLE2000, Corning Inc.). In detail, ultrashort pulses (220 nJ energy, $\sim$300~fs duration, 1030~nm wavelength) from a Yb:KYW cavity-dumped mode-locked oscillator are focused in the bulk of the glass using a 0.6~NA microscope objective. High-accuracy computer-controlled three-axis translators (Aerotech FIBERGlide) move the substrate at the constant speed of 40~mm s$^{-1}$ under the laser focus, allowing to draw the complex interferometric circuits in the three dimensions. The average depth below the glass surface of the inscribed circuits is 170~$\mu$m. Waveguides yield single-mode behaviour at 785~nm wavelength with propagation loss of about 0.5~dB cm$^{-1}$. The footprint of the 4- and 8- mode interferometers is respectively $\sim$ 25~mm~$\times$0.4~mm and $\sim$ 50~mm~$\times$0.9~mm.

\subsection*{Generation, manipulation and detection}

Single photons were generated at 785 nm with a type-II parametric down-conversion process pumping a crystal (2-mm long BBO) with a 392.5 nm wavelength Ti:Sa pulsed laser. The two photons are spectrally filtered by means of 3 nm interferential filters, and coupled into single mode fibers. The indistinguishability of the photons is then reached by a polarization compensation stage, and by propagation through delay lines for each possible path (used to adjust the degree of temporal distinguishability) before injection into the interferometer via a single mode fiber array. After the evolution through the integrated devices, photons are collected via a multimode fiber array. The detection system consists of 8 single-photon avalanche photodiodes used for the 8-mode chip.  8-channel electronic data acquisition system (ID800 by IDQuantique) allowed us to detect two-photon coincidences between all pairs of output states. LabView and C programs have been used to retrieve coincidences events associated to all possible combination of output modes. Two-photon bunching events from the same output mode have been collected connecting a 50:50 in-fiber beam splitter to each output multimode fiber.
